\documentclass[twocolumn, superscriptaddress]{aastex62}
\usepackage{amsmath}
\usepackage{txfonts}
\usepackage{tensor}
\usepackage{mathrsfs}
\usepackage[referable]{threeparttablex}
  
\newtheorem{assumption}{Assumption}
\renewcommand{\d}[0]{\mathrm{d}}

\renewcommand{\vec}[1]{\mathbf{#1}}
\newcommand{\df}[2]{\partial_{#2} {#1}}

\definecolor{ref}{rgb}{0.58, 0.0, 0.83}
\usepackage{tcolorbox}
\newtcolorbox{refereehelper}{colback=green!5!white,colframe=green!75!black}

\begin{document}
\author[0000-0002-6917-0214]{K.~S.~Croker}
\affiliation{Department of Physics and Astronomy, University of Hawai`i at M\=anoa,  2505 Correa Road, Honolulu, HI 96822, USA}

\author{J.~L.~Weiner}
\affiliation{Department of Mathematics, University of Hawai`i at M\=anoa, 2565 McCarthy Mall, Honolulu, HI 96822, USA}

\correspondingauthor{K.~S.~Croker}
\email{kcroker@phys.hawaii.edu}

\received{2019 February 12}
\revised{2019 June 8}
\accepted{2019 July 15}
\published{2019 August 28}

\title{Implications of Symmetry and Pressure in Friedmann Cosmology. I. Formalism}

\begin{abstract}
  We show that derivation of Friedmann's equations from the Einstein-Hilbert action, paying attention to the requirements of isotropy and homogeneity {during the variation}, leads to a different interpretation of pressure than what is typically adopted.
  Our derivation follows if we assume that the unapproximated metric and Einstein tensor have convergent perturbation series representations on a sufficiently large Robertson-Walker coordinate patch.
  We find the source necessarily averages all pressures, everywhere, including the interiors of compact objects.
  We demonstrate that our considerations apply (on appropriately restricted spacetime domains) to the {Kerr} solution, the Schwarzschild constant-density sphere, and the static de-Sitter sphere.
  {From conservation of stress-energy, it follows that material contributing to the averaged pressure must shift locally in energy.
  We show that these cosmological energy shifts are entirely negligible for non-relativistic material.
  In relativistic material, however, the effect can be significant.}
   We comment on the implications of this study for the dark energy problem.
\end{abstract}

\shortauthors{Croker and Weiner}
\shorttitle{Symmetry and pressure in Friedmann cosmology}


\section{Introduction}
\label{sec:intro}
\label{sec:metric_defn}
The foundations of modern cosmology, by \citet{peebles1980large}, \citet{bardeen1980gauge}, and \citet{kodama1984cosmological} have become standard textbook material \citep[e.g.][]{DodelsonCosmology, hu2004covariant}.
All these treatments begin from Einstein's equations 
\begin{align}
  G^{\mu\nu}(\eta, \vec{x}) = 8\pi GT^{\mu\nu}(\eta, \vec{x}).
  \label{eqn:typical_einstein}
\end{align}
One then assumes that a reasonable description of the cosmos is a perturbed spatially flat Robertson-Walker (RW) geometry 
\begin{align}
  g_{\mu\nu} &\equiv a^2(\eta)\left[\eta_{\mu\nu} + \epsilon h_{\mu\nu}^{(1)}(\eta, \vec{x})\right] \label{eqn:typical_metric},
\end{align}
where
\begin{align}
    \eta_{\mu\nu}\d x^\mu \d x^\nu &\equiv -\d \eta^2 + \d\vec{x}^2.
\end{align}
This metric encodes the Copernican Principle: on large scales, the universe has no preferred locations or directions.
In other words, the zero-order universe is homogeneous and isotropic.

Almost immediately, the procedure encounters trouble.
Substitution of Equation~(\ref{eqn:typical_metric}) into Equation~(\ref{eqn:typical_einstein}) gives 
\begin{align}
  G^{\mu\nu}(\eta) + O(\epsilon)(\eta, \vec{x}) = 8 \pi GT^{\mu\nu}(\eta, \vec{x}).
\end{align}
This equation is inconsistent if $T^{\mu\nu}$ encodes $O(1)$ position-dependent sources.
Gravitational radiation signatures of ultra-compact object mergers, as first reported by \citet{abbott2016observation}, provide direct evidence that such sources exist.
{To address this issue}, one can {instead} proceed with the following definition: 
\begin{align}
  T^{\mu\nu}(\eta, \vec{x}) \equiv T^{\mu\nu}_{(0)}(\eta) + \epsilon T^{\mu\nu}_{(1)}(\eta, \vec{x}).
  \label{eqn:artificial_stress}
\end{align}
Isotropy and homogeneity then constrain the $O(1)$ contribution into the form of a perfect fluid 
\begin{align}
  T^{\mu\nu}_{(0)} = \left[\rho(\eta) + \mathcal{P}(\eta)\right]u^\mu u^\nu + \mathcal{P}(\eta)a^{-2}\eta^{\mu\nu}.
  \label{eqn:trouble_end}
\end{align}
The functions $\mathcal{P}(\eta)$ and $\rho(\eta)$ can then be determined from data, and compared to theoretical expectations.

To predict $\mathcal{P}(\eta)$ and $\rho(\eta)$ in the {redshift $z \gtrsim 40$} universe, overwhelming evidence of a hot big bang justifies summing the second moments of a Boltzmann distribution $f_i$ for each species $i$ at some temperature $T_i$ 
\begin{align}
  T^{\mu\nu}_{(0)}(\eta) \propto \sum_i \int \frac{p^\mu p^\nu}{p^0} f_i\left[|\vec{p}|, T_i(\eta)\right]~\d^3 p.
\end{align}
In the $z \lesssim 40$ universe, however, structures {may have begun to} form and appropriate distribution functions are unknown.
In the spirit of \citet[][\S3]{kaiser2017there}, we can try to estimate $\mathcal{P}(\eta)$ and $\rho(\eta)$ at late times with linear perturbation theory.
We begin from a single star: 
\begin{enumerate}
\item{\textbf{Hypothesis (vacuum)}: exterior to the star is {effectively} vacuum.}
\item{\textbf{Hypothesis (spherical symmetry)}: the star is spherical to a very good approximation.}
\item{$\therefore$ \textbf{Birkhoff's theorem} $\implies$ the star is perceived as a point mass to good approximation.}
\end{enumerate}
At sufficient distance from the star, the Newtonian monopole limit is recovered.
One then invokes superposition and averages over an ensemble of such point masses to produce $\rho(\eta)$.
{ Performing these averages over {distinct} regions in space reveals a lower-bound volume $\mathcal{V}$, beyond which the $\rho(\eta)$ computed at different points agree.}
{The observational success of linear perturbation theory} supports the expectation that the rest-mass densities of such ensembles should dominate any kinetic contributions by $\sim 10^5$.
Alternatively, following \citet[][p.~474]{WeinbergGR} or \citet[][p.~297]{peebles1993principles}, one applies Birkhoff to a round ball\footnote{in geometer language, a ``ball'' is anything diffeomorphic to a ``round ball.''} cut out from some presumed RW universe.
The interior region is now vacuum and the conclusion becomes that whatever was inside the region can be averaged and reintroduced as a point mass.
In this case, the interior motions contribute zero pressure to the exterior universe.
Either way, the predicted contribution to $T_{(0)}^{\mu\nu}(\eta)$ is a $\mathcal{P}(\eta) \simeq 0$ perfect fluid.
{ Observations, however, are consistent with a late-time universe dominated instead by an apparently fixed-density $\mathcal{P}(\eta)=-\rho(\eta)$ perfect fluid.}

{ At this point, we have built a cosmological model by assuming an RW metric ansatz and then constructing an appropriate stress tensor.
One could instead imagine starting with Einstein's equations for the actual matter distribution and ask what must be done to arrive at an RW metric.
This alternative approach proceeds as follows, where we summarize aspects of \citet[][\S1]{wetterich2003can} and \citet[][\S2]{ellis2011inhomogeneity}.
Observe that the matter in the visible universe is distributed uniformly on sufficiently large spatial volumes $\mathcal{V}$.
This suggests taking a spatial average of Einstein's Eqns.~(\ref{eqn:typical_einstein}) 
\begin{align}
  \left<G^{\mu\nu}(\eta, \vec{x})\right>_\mathcal{V} = 8\pi G \left<T^{\mu\nu}(\eta, \vec{x})\right>_\mathcal{V}, \label{eqn:backreaction-einstein}
\end{align}
and defining
\begin{align}
T^{\mu\nu}_{(0)}(\eta) \equiv \left<T^{\mu\nu}(\eta, \vec{x})\right>_\mathcal{V}.
\end{align}
The objects appearing in this equation are components, so care must be taken to ensure that the averages are covariant.
Suppose we perform averages with an RW metric defined\footnote{{ As pointed out by \citet[][see the text below Equation~(7)]{wetterich2003can}, constructing an RW metric via averaging may lead to a metric significantly different than that inferred from dynamical measurements within the actual spacetime.}} by
\begin{align}
g_{\mu\nu}^{\mathrm{RW}}(\eta) \equiv \left<g_{\mu\nu}(\eta, \vec{x})\right>_\mathcal{V}.
\end{align}
This is reasonable because a spatially averaged quantity is isotropic and homogeneous, by construction.
At early times, we expect $g_{\mu\nu} \simeq g^\mathrm{RW}_{\mu\nu}$ because gravitational collapse has not yet introduced spatial dependence into the matter distribution.
The nonlinearity of the Einstein tensor as a function of the metric, however, implies that 
\begin{align}
      G^{\mu\nu}(\eta) \equiv G^{\mu\nu}\left[g_{\mu\nu}^\mathrm{RW}\right] \neq \left<G^{\mu\nu}\left[g_{\mu\nu}(\eta, \vec{x})\right]\right>_\mathcal{V}.
\end{align}
in general.
Here we have used square braces to indicate that $G^{\mu\nu}$ is a differential operator acting on the metric.
The discrepancy is expected to grow larger as gravitational collapse forms inhomogeneous, small-scale structures.
If one defines the difference
\begin{align}
    X^{\mu\nu}(\eta) \equiv G^{\mu\nu}(\eta) - \left<G^{\mu\nu}\left[g_{\mu\nu}(\eta, \vec{x})\right]\right>_\mathcal{V},
\end{align}
then the spatially averaged Einstein's Eqns.~(\ref{eqn:backreaction-einstein}) become
\begin{align}
G^{\mu\nu}(\eta) = 8\pi G T^{\mu\nu}_{(0)}(\eta) + X^{\mu\nu}(\eta).
\end{align}
This additional term $X^{\mu\nu}$ is called the cosmological backreaction.
It has been claimed by \citet{rasanen2004dark}, \citet{kolb2006cosmic}, and others that the cosmological backreaction will manifest as an apparent $\mathcal{P}(\eta) = -\rho(\eta)$ perfect fluid.
This purported ability to explain the accelerated late-time expansion of the universe, by averaging the left side of Einstein's equations, has generated significant interest in the backreaction community.

There is vigorous disagreement, most recently between \citet{green2014well} and \citet{buchert2015there}, about the relevance of cosmological backreaction.
We believe that this disagreement can be distilled into a single question: what is the correct way to construct a large-scale homogeneous model of a small-scale inhomogeneous universe within the framework of GR? 
It is clear that construction of any isotropic and homogeneous global model from Einstein's local differential equations requires removing position dependence.
Indeed, both the Birkhoff approach and the backreaction approach take spatial averages. 
The lack of any unique prescription for which averages to perform, however, is a fundamental ambiguity in our understanding of how to use Einstein's equations to extract physics on the largest scales.

In this paper, we show how to resolve the ambiguity.}
We develop an internally consistent perturbative treatment of both {the} metric \emph{and} {the} stress tensor directly from the Einstein-Hilbert (EH) action.
The main results of this paper are as follows:
\begin{itemize}
{ \item{Friedmann's acceleration equation and conservation of stress energy without effective source terms of geometric origin, together with;}}
\item{an explicit expression for $\mathcal{P}(\eta)$, which may not vanish at late times.}
\end{itemize}
{ These results are to be contrasted with expectations, respectively, of an additional backreaction term and a pressure-free stress tensor via Birkhoff's theorem.}

The rest of this paper is organized as follows.
In \S\ref{sec:zero-order}, we precisely define our gravitational model within the context of GR.
We then derive the appropriate equations of motion, taking great care to enforce the RW model symmetries at each step of the calculation.
In \S\ref{sec:strong}, we determine how various astrophysical sources, including strong ones, contribute to the equations of motion.
{In \S\ref{sec:consequences}, we highlight the observational consequences of $\mathcal{P}(\eta)\neq 0$.}
In \S\ref{sec:conclusion}, we summarize our results.
In Appendix~\ref{sec:boundary_term}, we provide a detailed discussion of boundary terms.
In Appendix~\ref{sec:constant_density_einstein}, we provide additional calculations, which support \S\ref{sec:schwarzschild_interior}.
Throughout this paper, we set $c \equiv h \equiv 1$ and use the $(-,+,+,+)$ signature.

\section{Friedmann's equations from the action}
\label{sec:zero-order}
We adopt GR as encoded in the EH action.
Everywhere on the manifold $\mathcal{M}$, there exists a well-defined\footnote{{i.e. unique and representation independent}} metric $\mathbf{g}$ and a well-defined collection of other fields $\mathbf{\Psi}$.
These combine to give a well-defined stress-tensor $\mathbf{T}$.
We note that the model builder's arbitrary choice for approximating $\mathbf{g}$ does not change the actual $\mathbf{\Psi}$.
We make these statements precise, beginning with the following 
\begin{assumption}
  There exists a series representation (this includes the coordinates) for the actual metric 
  \begin{align}
    g_{\mu\nu}(\eta, \vec{x})\Bigg|_{\mathcal{U}\subset \mathcal{M}} \equiv \lim_{N\to\infty}a(\eta)^2\left[\eta_{\mu\nu} + \sum_{n = 1}^N \epsilon^n h^{(n)}_{\mu\nu}(\eta, \vec{x})\right] \qquad \epsilon < 1 \label{eqn:metric_picture_defn}, 
  \end{align}
  which converges on some compact submanifold (with boundary) $\mathcal{U} \subset \mathcal{M}$.
  The components $h_{\mu\nu}^{(n)}$ are at least twice continuously differentiable $(C^2)$ on $\mathcal{U}$.
\end{assumption}%
This representation need not be unique.
{For the purposes of all subsequent discussions, however, we only require the existence of such convergent representations.}
If $\mathcal{U}$ is not connected, then we take Assumption 1 to mean that there exists such a representation on each connected component.
The factors of $\epsilon^n$ encode the relative magnitude of the terms in the expansion.
In other words, for fixed $\mu$ and $\nu$, each component $-1 \leqslant h_{\mu\nu}^{(n)} \leqslant 1$ for all $n$.
This guarantees that the representation is dominated by a convergent power series.
{Truncation of Equation~(\ref{eqn:metric_picture_defn}) at $N=1$ produces Equation~(\ref{eqn:typical_metric}).
  In this sense, Equation~(\ref{eqn:metric_picture_defn}) generalizes the covariant linear perturbation theory metric ansatz to arbitrary order.}

Formal substitution of Equation~(\ref{eqn:metric_picture_defn}) into Einstein's equations will give a formal power series in $\epsilon$.
We do not know, \emph{a priori}, the relative magnitudes of these terms or if this series converges.
This occurs because the Einstein tensor and the stress tensor are proportional and we have not constrained the matter fields $\mathbf{\Psi}$.
Since we wish to solve Einstein's equations {perturbatively,} order-by-order in $\epsilon$ on $\mathcal{U}$, we make the following
\begin{assumption}
  {There exist length scales $L_n(\eta, \vec{x})$ such that} the Einstein tensor can be expanded as 
  \begin{align}
  G_{\mu\nu}(\eta, \vec{x})\Bigg|_{\mathcal{U}\subset \mathcal{M}} \equiv \lim_{N\to\infty} \sum_{n = 0}^N \epsilon^n \left(\frac{\bar{G}^{(n)}_{\mu\nu}}{L_n^2}\right) (\eta, \vec{x}) \qquad \epsilon < 1 \label{eqn:einstein_picture_defn},
  \end{align}
  where $\left|\bar{G}_{\mu\nu}^{(n)}\right| \leqslant 1$ and $\epsilon$ is the same as in Equation~(\ref{eqn:metric_picture_defn}).
\end{assumption}
{Following \citet[\S6.3]{lin1988mathematics}, these length scales are combined, at each order, with length scales set by derivatives of $\mathbf{\Psi}$ within the stress tensor to produce dimensionless field equations.
The existence of $L_n$ is} reasonable because the Einstein tensor contains at most second derivatives of the metric, which we have defined to be at least $C^2$.
The components of the Einstein tensor are thus continuous on a compact set, and thus bounded.
{Note that $L_n \neq 0$, as we have assumed that Einstein's equations are well-defined everywhere on $\mathcal{M}$, and $\mathcal{U} \subset \mathcal{M}$.
    The particular finite values of $L_n$ do not matter.
    Assumption 2 guarantees their existence so that field equations, at each order in the perturbation, can be well-defined.}

Taken together, Assumptions 1 and 2 {establish sufficient conditions for} Einstein's equations to be solved perturbatively on $\mathcal{U}$.
They do not assert that low-order approximations are always useful.
On the contrary, it may be that infinitely many terms are required to describe a region of spacetime.
Our concern is not with utility, but with existence of well-defined representations for $\mathbf{g}$.
This will guarantee that equations governing the lowest-order terms remain valid even on submanifolds of $\mathcal{U}$ with strong gravity.
{As is standard in the literature, we will indicate the order $N$ at which field equations are truncated with $O\left(\epsilon^N\right)$, even though we write dimensionful field equations.
No confusion will arise, as Assumptions 1 and 2 guarantee that a perturbative solution is always possible.}

\subsection{The Action Principle for Gravity}
{ Given Assumptions 1 and 2, we work exclusively on $\mathcal{U}$.}
In the context of gravity, the action principle demands that 
\begin{align}
  \delta\left\{S_G\left[g_{\mu\nu}\right] + S_M\left[g_{\mu\nu}, \mathbf{\Psi}(\eta, \vec{x}) \right]\right\}\Bigg|_{\mathbf{\Psi}} \equiv 0 \label{eqn:action-principle}
\end{align}
where $S_G$ is a gravitational action, $S_M$ is a matter action, and $\mathbf{\Psi}$ is held fixed for the variation {$\delta$}.
By the definition given in \citet[][\S12.2]{WeinbergGR}, the symmetric, rank-(2,0), stress tensor $T^{\mu\nu}$ enters Einstein's equations through
\begin{align}
  \delta S_M\Bigg|_{\mathbf{\Psi}} \equiv \frac{1}{2}\int_\mathcal{U} T^{\mu\nu}\delta g_{\mu\nu}\sqrt{-g}~\d^4x.
  \label{eqn:hilbert_stress}
\end{align}
Here, $g$ is the metric determinant.
Note that we have flipped indices, relative to Weinberg's definition, and thus incur a minus sign through the metric variation 
\begin{align}
  \delta\left( \tensor{\delta}{^\alpha_\beta} \right) = 0 = \delta\left(g^{\alpha\mu}g_{\beta\mu}\right).
\end{align}
Weinberg's definition encodes the standard assumption that $S_M$ comes from a Lagrange scalar density $\mathscr{L}_M$,
  \begin{align}
    \delta S_M &= \int_\mathcal{U} \delta \left(\mathscr{L}_M \sqrt{-g}\right)~\d^4 x \\
    &= \int_\mathcal{U} \left[\frac{1}{\sqrt{-g}}\frac{\delta}{\delta g_{\mu\nu}}\left(\mathscr{L}_M\sqrt{-g}\right)\right]\delta g_{\mu\nu} \sqrt{-g}~\d^4 x. \label{eqn:varied_matter_action}
  \end{align}
{Comparison of Equation~(\ref{eqn:hilbert_stress}) with Equation~(\ref{eqn:varied_matter_action}) reveals} that, \emph{under the integral}, one may identify 
\begin{align}
  T^{\mu\nu} =  \frac{2}{\sqrt{-g}}\frac{\delta}{\delta g_{\mu\nu}} \left\{\mathscr{L}_M\left[\mathbf{\Psi}(\eta, \vec{x}), g_{\mu\nu}\right]\sqrt{-g}\right\}.
  \label{eqn:integrand_stress_tensor}
\end{align}
The stress tensor components are, therefore, a mixture of the metric and the non-gravitational contributions $\mathbf{\Psi}(\eta, \vec{x})$.
{The non-gravitational contributions $\mathbf{\Psi}(\eta, \vec{x})$ have no explicit dependence on $\mathbf{g}$, and thus are unaltered by the model builder's choice of metric model.}
Of course, the $\mathbf{\Psi}(\eta, \vec{x})$ are implicitly constrained by the eventual equations of motion.
Any premature constraint, before equations of motion consistent with the action principle are determined, risks introducing inconsistencies into the field equations.

Let $R$ be the Ricci scalar, built from the Levi-Civita connection of any metric.
The EH gravitational action is 
\begin{align}
  S_G &\equiv \frac{1}{16\pi G} \int_\mathcal{U} R\sqrt{-g}~\d^4x.
  \label{eqn:eh-action}
\end{align}
We use the representation of the metric given in Equation~(\ref{eqn:metric_picture_defn}).
{ For the purposes of computation, we now parameterize $\mathcal{U}$.}
{While there is} observational evidence that the universe began in a nearly singular hot and dense point, it is not clear how far back in time one can ``safely'' apply GR.
We thus restrict consideration to 
\begin{align}
  \eta \in \left[\eta_i, \eta_f\right].  
\end{align}
Here $\eta_i$ is some arbitrary initial time and $\eta_f > \eta_i$ can either be arbitrary, in the case of matter domination, or any time below the asymptotic value in the case of dark energy domination.
Note that because $\mathcal{U}$ is compact, an action defined over it will never diverge.
For the moment, consider a closed 3-ball $\mathcal{V} \subset \mathbb{R}^3$ so that 
\begin{align}
  \mathcal{U} \equiv \mathcal{V} \times \left[\eta_i, \eta_f\right] \label{eqn:manifold}.
\end{align}
We will often use $\mathcal{V}$ to represent both this compact submanifold and its volume, which is the integral of the Euclidean volume form over this submanifold.

We now work only through $O(1)$.
From Equation~(\ref{eqn:metric_picture_defn}), to leading order, the approximation is 
\begin{align}
  g_{\mu\nu} = a^2(\eta)\left[\eta_{\mu\nu} + O(\epsilon)\right].
\end{align}
This gives, for the metric determinant and inverse metric variation,
\begin{align}
  \sqrt{-g} &= a^4 + O(\epsilon) \\
  \delta g^{\mu\nu} &= -2\eta^{\mu\nu}a^{-3} \delta a  + O(\epsilon).
\end{align}
Following \citet[Equation~(D.9), p.~446]{WaldGR}, we may exploit that the flat RW metric is a conformal rescaling of flat space to write 
\begin{align}
  R = -6\eta^{\mu\nu}a^{-3} \partial_\mu\partial_\nu a + O(\epsilon), 
\end{align}
where the derivatives are simple partials.
Note that we have used Assumption 2 to regard the Ricci scalar of the bare (not multiplied by the scale factor) metric as $O(\epsilon)$.

Substitution into the definition of the gravitational action $S_G$ gives
\begin{align}
  S_G = -\frac{3}{8\pi G} \int_\mathcal{U} \eta^{\mu\nu} a(\partial_\mu\partial_\nu a)~\d^4x + O(\epsilon).
\end{align}
Performing the variation gives 
\begin{align}
  \delta S_G = -\frac{3}{8\pi G} \int_\mathcal{U} \left[ \delta a(\partial^\mu\partial_\mu a) + a(\partial^\mu\partial_\mu \delta a) \right]\d^4x 
\end{align}
plus terms $O(\epsilon)$.
Integrating by parts twice gives 
\begin{align}
  \delta S_G = -\frac{3}{4\pi G} \int_\mathcal{U} \delta a \partial^\mu\partial_\mu a~\d^4 x + \mathcal{B} + O(\epsilon) \label{eqn:almost_there}
\end{align}
where we have separated off a total divergence term 
\begin{align}
  \mathcal{B} = -\frac{3}{8\pi G} \int_\mathcal{U} \partial_\mu \left\{\eta^{\mu\nu}\left[a \partial_\nu\delta a - \delta a \partial_\nu a \right]\right\}\d^4x. \label{eqn:boundary}
\end{align}
This term becomes a boundary term, which is usually discarded.
We must be more careful in our RW setting, however, because we do not have spatial control of the variations $\delta a(\eta)$.
Detailed consideration of the boundary term can be found in Appendix~\ref{sec:boundary_term}, where we show that the term should still be discarded.
Consistent with this result, we now set $\mathcal{B} = 0$.

\subsection{Symmetry of the Variation $\delta a$}
We now arrive at the crucial step, which reveals the importance of working directly from the EH action.
Only variations consistent with the RW model symmetries are permitted.
In other words, the variations $\delta a$ of the scale factor $a(\eta)$ must be functions of conformal time alone: $\delta a(\eta)$.
Since the integrand in Equation~(\ref{eqn:almost_there}) depends only on $\eta$, we have that
\begin{align}
  \delta S_G = -\frac{3}{4\pi G}\int_{\eta_i}^{\eta_f} \delta a \partial^\mu\partial_\mu a~\d\eta \int_\mathcal{V}~\d^3x + O(\epsilon).
\end{align}
Here we have used Fubini's theorem to write the compactly supported integral as an iterated integral.
Note that the spatial integration is just the { comoving} volume $\mathcal{V}$, so 
\begin{align}
  \delta S_G = -\frac{3}{4\pi G}\int_{\eta_i}^{\eta_f} \delta a \partial^\mu\partial_\mu a~\mathcal{V} \d\eta + O(\epsilon).
  \label{eqn:zero-order-grav-variation}
\end{align}

To compute the variation of the matter action, first note that 
\begin{align}
  T^{\mu\nu}\delta g_{\mu\nu} &= 2a\eta_{\mu\nu}T^{\mu\nu}\delta a  + O(\epsilon) \\
  &= 2a^{-1}\delta a \left[g_{\mu\nu} - O(\epsilon)\right]T^{\mu\nu} + O(\epsilon) \\
  &= 2a^{-1}\tensor{T}{^\mu_\mu}\delta a + O(\epsilon) 
\end{align}
where $\tensor{T}{^\mu_\mu}$ is the trace of the stress tensor.  
Substitution into Equation~(\ref{eqn:hilbert_stress}), restricted to $\mathcal{U}$, gives 
\begin{align}
  \delta S_M = \int_{\eta_i}^{\eta_f} \int_\mathcal{V} a^3 \tensor{T}{^\mu_\mu}(\eta, \vec{x}) \delta a ~\d^3x~\d\eta + O(\epsilon).
\end{align}
Again, because $a$ and $\delta a$ depend only on $\eta$, we have that
\begin{align}
  \delta S_M = \int_{\eta_i}^{\eta_f} a^3 \delta a \left [\int_\mathcal{V} \tensor{T}{^\mu_\mu}(\eta, \vec{x}) ~\d^3x\right]~\d\eta + O(\epsilon).
  \label{eqn:zero-order-matter-variation}
\end{align}
The trace of the stress tensor cannot be moved through the integral. 
{The trace continues to be position-dependent, even though the metric at zero order is position-independent, because the stress tensor depends on $\mathbf{\Psi}(\eta, \vec{x})$.}
These fields are not explicitly constrained by our (or any) choice of metric model.
The quantity in the square brackets, because of the integration over $\mathcal{V}$, is formally independent of position.
Note, however, that it could be sensitive to the choice of the region $\mathcal{V}$.
This will be elaborated upon in \S\ref{sec:V}, where we discuss the physical significance of $\mathcal{V}$.

We now combine the variation of the gravitational action Equation~(\ref{eqn:zero-order-grav-variation}) with that of the matter action Equation~(\ref{eqn:zero-order-matter-variation}).
The action principle as stated in Equation~(\ref{eqn:action-principle}) {becomes}
\begin{align}
  \int_{\eta_i}^{\eta_f} \delta a \left[ -\frac{3}{4\pi G} \partial^\mu\partial_\mu a~\mathcal{V} + a^3 \int_\mathcal{V}  \tensor{T}{^\mu_\mu}(\eta, \vec{x})~\d^3 x \right] \d\eta \equiv 0.
\end{align}
{Applying} the Fundamental Theorem of Variational Calculus {then gives} the consistent equation of motion 
\begin{align}
  \frac{3}{4\pi G} \partial^\mu\partial_\mu a~\mathcal{V} = a^3 \int_\mathcal{V}  \tensor{T}{^\mu_\mu}(\eta, \vec{x})~\d^3 x.
\end{align}
Dividing by the constants on the left naturally reveals 
\begin{align}
  \partial^\mu\partial_\mu a &= \frac{4 \pi G}{3} a^3 \frac{1}{\mathcal{V}} \int_\mathcal{V}  \tensor{T}{^\mu_\mu}(\eta, \vec{x})~\d^3 x \\
  &= \frac{4 \pi G}{3} a^3 \left<\tensor{T}{^\mu_\mu}(\eta, \vec{x})\right>_\mathcal{V}, 
\end{align}
which is the spatial-slice average of the stress tensor's trace.
{ This expression is manifestly coordinate-invariant:
\begin{itemize}
\item{the trace of any tensor is a general coordinate scalar; and}
\item{by Assumption 1, the notion of a spatial slice is geometrically well-defined \citep[e.g.][\S12, p.~342]{DaddyONeill}}
 \end{itemize}
The spatial integration in this average, and the volume to normalize it, come directly from the action principle.}
Expanding the derivatives and trace gives Friedmann's acceleration equation 
\begin{align}
  \frac{\d^2 a}{\d \eta^2} = \frac{4\pi G}{3} a^3\left<\rho(\eta, \vec{x}) - \sum_{i=1}^3\mathcal{P}_i(\eta, \vec{x})\right>_\mathcal{V} \label{eqn:friedmanns_equation}
\end{align}
expressed in conformal time.
Though the trace is invariant, $\rho(\eta, \vec{x})$ and $\mathcal{P}_i(\eta, \vec{x})$ are the diagonal components of the generic stress tensor in the RW preferred coordinate system.
This is the coordinate system in which the metric takes the form of Equation~(\ref{eqn:metric_picture_defn}).
One may introduce the definitions 
\begin{align}
 \rho(\eta)  &\equiv  \big<\rho(\eta, \vec{x})\big>_\mathcal{V}  \label{eqn:friedmann_rho}\\
   3\mathcal{P}(\eta) &\equiv \left<\sum_{i=1}^3\mathcal{P}_i(\eta, \vec{x})\right>_\mathcal{V} \label{eqn:friedmann_P}  
\end{align}
to arrive at the typical presentation of Friedmann's acceleration equation in conformal time.
Note, however, that these defined $\rho(\eta)$ and $\mathcal{P}(\eta)$ must be spatial-slice averages of the unapproximated quantities.

\subsection{Significance of the Submanifold $\mathcal{V}$}
\label{sec:V}
The spatial average contains the arbitrary closed 3-ball $\mathcal{V}$.
This enters the construction because we must guarantee that the action integral remains well-defined.
This means that Equation~(\ref{eqn:friedmanns_equation}) represents a continuum of Friedmann models.

We may understand the parameter $\mathcal{V}$ as follows. 
Fix a radius $b$ and pick some event $P \in \mathcal{U}$.
Compute the spatial averages, at all $\eta$, of the stress trace over a round ball of radius $b$ centered at the spatial location of $P$.  
The resulting source will {determine how $a(\eta)$ evolves in conformal time}. 
With the same $b$, repeat this procedure at a distinct event $Q \in \mathcal{U}$, which lies in the same spatial slice as $P$.
If the $Q$ spatial averages do not agree with the $P$ spatial averages, the dynamics for $a(\eta)$ need not agree.
This can happen because matter is not distributed uniformly on small scales.
In other words, well-defined field equations for $a(\eta)$ require a volume large enough so that the averaged density and pressures are position-independent.

{ Observations \citep[e.g.][]{hogg2005cosmic, scrimgeour2012wigglez, nadathur2013seeing} suggest} that, at the present epoch, there is a radius
\begin{align}
  b_* \sim {180}~\mathrm{Mpc},
\end{align}
beyond which the averaged quantities are position-independent.
We will assume that $\mathcal{V}$ contains a round ball of at least radius $b_*$.
{ Note that the specific value of $b_*$ is unimportant for the purposes of our argument.}
Strong evidence for hierarchical formation of structure tells us that, throughout time, these averaged quantities remain position-independent.
In other words, { any such} $\mathcal{V}$ suffices for all earlier epochs.

{ 
\subsection{Resolution of the Averaging Ambiguity}
Friedmann's equation, as given in Equation~(\ref{eqn:friedmanns_equation}) with $\mathcal{V}$ interpreted according to \S\ref{sec:V}, resolves all ambiguity in our understanding of how to use Einstein's equations to extract physics on the largest scales.  
Consistent with our motivation of cosmological backreaction in \S\ref{sec:intro}, the action includes an integration of degrees of freedom present in the metric over the spatial volume $\mathcal{V}$.
At zero order, however, the only such degree of freedom is the scale factor $a(\eta)$.
It is already constrained by the RW model assumption (not the universe) to be position-independent.
The required integration produces the spatial-slice volume $\mathcal{V}$, as seen in Equation~(\ref{eqn:zero-order-grav-variation}).
In other words, given Assumptions 1 and 2, the action principle produces equations of motion without any cosmological backreaction $X^{\mu\nu}(\eta)$.}

{It is important to note that, given a different starting ansatz, one can arrive at a different cosmological model.
    This different cosmological model could possibly exhibit a backreaction: additive and position-independent contributions of geometric origin.
    The presence or absence of such terms is an artifact of the particular model and is logically distinct from the actual metric tensor on $\mathcal{U}$.
    Our Assumption 1 reduces to the RW metric when truncated at $N=0$, and to the typical covariant linear perturbation metric when truncated at $N=1$.
    What we have shown is that, within these frameworks, there is zero cosmological backreaction $X^{\mu\nu}(\eta)$.
}

{
  {Inspection of Equation~(\ref{eqn:friedmanns_equation}) and Equation~(\ref{eqn:friedmann_P}) reveals that the position-independent pressure $\mathcal{P}(\eta)$ of any Friedmann model must include all contributions from pressures interior to compact objects on $\mathcal{V}$.}
In other words, given Assumptions 1 and 2, using Birkhoff's theorem to construct a cosmological source to Friedmann's equations is inconsistent with the action principle.
In general, use of Birkhoff's theorem in cosmological contexts is only appropriate under very restricted settings, which we derive and discuss in \S\ref{sec:birkhoff_oops}.
We summarize these statements for clarity.}
\begin{refereehelper}
  { 
Our Friedmann's equation is an unambiguous and coordinate-invariant consequence of Assumptions 1, 2, and the EH action.
\begin{itemize}
\item{There is no backreaction.  Inhomogenieties in the spatial distribution of matter do not affect the source to Friedmann's equation.}
\item{Pressures interior to all compact objects contribute to Friedmann's equation.}
\end{itemize}
These consequences {arise} because the RW metric cannot distinguish spatial regions.
Perturbatively consistent equations of motion thus require a zero order source without explicit or \emph{implicit} notions of interior.}
\end{refereehelper}

\subsection{Covariant Conservation of Stress-Energy at Zero Order}
\label{sec:zero-order_conservation}
{ It is instructive to show how careful application of symmetry considerations also produces the consistent conservation of stress-energy statement.}
To derive the appropriate relation, note that a general coordinate scalar, such as the matter action, cannot change under coordinate transformations.
It is shown in \citet[\S12.3]{WeinbergGR} that this restriction implies 
\begin{align}
  \Delta S_M = 0 = \int_\mathcal{U} \epsilon^\lambda\nabla_\nu\tensor{T}{^\nu_\lambda}(\eta, \vec{x})\sqrt{-g}~\d^4x, \label{eqn:action_stmt_conservation}
\end{align}
where $\epsilon^\lambda$ is an arbitrary (infinitesimal) vector field.
Since $\epsilon^\lambda$ is arbitrary, the above equation immediately leads to the familiar statement of covariant conservation of stress-energy:
\begin{align}
  \nabla_\nu\tensor{T}{^\nu_\lambda}(\eta, \vec{x}) = 0.
\end{align}
Note that we deviate slightly from Weinberg's notation and use $\Delta$ to distinguish the change in $S_M$ induced by a coordinate change from any hypothetical change induced by the variational differential $\delta$.
We continue to interpret $\mathcal{V}$ consistent with \S\ref{sec:V}
For clarity, we will continue to explicitly write the time and position dependence of the generic stress tensor.

At zero order, the position-independent vector fields $\epsilon^\lambda(\eta)$ are the only ones permissible by RW model symmetries.
For consistency, we must restrict ourselves to these vector fields.
Note that Equation~(\ref{eqn:action_stmt_conservation}) is true for arbitrary $\epsilon^\lambda$, so it is true for position-independent $\epsilon^\lambda(\eta)$.
Thus, we may commute though the spatial integral as before:
\begin{align}
  \begin{split}
    0 = \int_{\eta_i}^{\eta_f} \epsilon^\lambda(\eta)a^4\int_\mathcal{V} \Big[&\df{\tensor{T}{^\nu_\lambda}(\eta, \vec{x})}{\nu} + \tensor{\Gamma}{^\nu_{\nu\rho}}\tensor{T}{^\rho_\lambda}(\eta, \vec{x}) \\
      &- \tensor{\Gamma}{^\rho_{\lambda\nu}}\tensor{T}{^\nu_\rho}(\eta, \vec{x})\Big]~\d^3x~\d\eta + O(\epsilon).
  \end{split} \label{eqn:conservation-volume}
\end{align}
This is permissible because the preferred RW frame is a function of $\eta$ alone.
We now expand the derivative term and apply Stokes' theorem on the spatial slice 
\begin{align}
  \begin{split}
    \int_{\eta_i}^{\eta_f} \epsilon^\lambda(\eta)a^4\Bigg\{\int_\mathcal{V} \Big[&\df{\tensor{T}{^0_\lambda}(\eta, \vec{x})}{0} + \tensor{\Gamma}{^\nu_{\nu\rho}}\tensor{T}{^\rho_\lambda}(\eta, \vec{x}) \\
      &- \tensor{\Gamma}{^\rho_{\lambda\nu}}\tensor{T}{^\nu_\rho}(\eta, \vec{x})\Big]~\d^3x \\
    &+ \int_{\partial\mathcal{V}} \tensor{T}{^k_\lambda}(\eta, \vec{x})~\d^2x_k\Bigg\}~\d\eta = 0. 
  \end{split} \label{eqn:conservation_boundary}
\end{align}
Because we are interested in the conservation condition, set $\lambda=0$ so that the vector $\tensor{T}{^k_0}$ is an energy flow.
By Einstein's equations $\tensor{T}{^k_0} \sim \tensor{G}{^k_0}$, but $\tensor{G}{^k_0} \sim O(\epsilon)$ by Assumption 2, so the boundary term must be dropped at this order
\begin{align}
  \begin{split}
    0 = \int_{\eta_i}^{\eta_f} a^4 \epsilon^0 \int_\mathcal{V} \Big[&\df{\tensor{T}{^0_0}(\eta, \vec{x})}{0} + \tensor{\Gamma}{^\nu_{\nu\rho}}\tensor{T}{^\rho_0}(\eta, \vec{x}) \\
      &- \tensor{\Gamma}{^\rho_{0\nu}}\tensor{T}{^\nu_\rho}(\eta, \vec{x})\Big]~\d^3x~\d\eta.  
  \end{split}
\end{align}
A standard calculation gives 
\begin{align}
  \tensor{\Gamma}{^0_{0\nu}} = H_\eta \tensor{\delta}{^0_\nu} \qquad \tensor{\Gamma}{^k_{0\nu}} = H_\eta\tensor{\delta}{^k_\nu} \qquad \tensor{\Gamma}{^\nu_\nu_\rho} = 4H_\eta\tensor{\delta}{^0_\rho}
\end{align}
where $H_\eta \equiv a^{-1}\d a/\d\eta$.
All of these connection components are dependent on $\eta$ alone, so that dividing through by $\mathcal{V}$ gives 
\begin{align}
  \frac{\partial}{\partial \eta}\left<\tensor{T}{^0_0}(\eta, \vec{x})\right>_\mathcal{V} + 3H_\eta\left<\tensor{T}{^0_0}(\eta, \vec{x})\right>_\mathcal{V} - H_\eta\left<\tensor{T}{^k_k}(\eta, \vec{x})\right>_\mathcal{V} = 0, \label{eqn:zero-order-conservation}
\end{align}
which is the appropriate continuum of conservation of stress-energy statements.

\subsection{Generalizations}
The most general RW metric takes the form 
\begin{align}
  \d s^2 = -b(\tau)^2\d \tau^2 + a(\tau)^2\d x^i \d x^j \gamma_{ij},
\end{align}
where $\gamma_{ij}$ can be the metric for any 3-space of constant curvature and $b(\tau)$ is a gauge degree of freedom.
In the previous computation, we have fixed the gauge by defining 
\begin{align}
  b(\tau)~\d \tau \equiv a(\eta)~\d \eta. 
\end{align}
If one keeps the $b(\tau)$ degree of freedom, and varies with respect to it, the typical Friedmann energy equation is obtained \citep[see][Equation~(9)]{suzuki1996classically}.
{In this way, one can reproduce Equation~(\ref{eqn:zero-order-conservation}) entirely within the variational formalism.}
Non-flat spatial slices lead to the same conclusions with respect to computation of the source terms.

The formalism developed in this section has also been applied to the next order in $\epsilon$.
The advantage is that source terms to the field equations are generated unambiguously.
This has been verified by \citet[][\S3]{CrokerThesis2018} at $O(\epsilon)$ for the scalar modes, where a lengthy calculation in longitudinal gauge reproduces the standard equations.

\section{Applicability of the metric representation to physical sources}
\label{sec:strong}
In the previous section, we showed that if Eqns.~(\ref{eqn:metric_picture_defn}) and (\ref{eqn:einstein_picture_defn}) converge on some region $\mathcal{U}$, then all densities and pressures in $\mathcal{U}$ will affect $a(\eta)$ according to Equation~(\ref{eqn:friedmanns_equation}).
In this section, we show that Equation~(\ref{eqn:friedmanns_equation}) continues to hold {on small scales and inside particular strong sources.}

We will proceed in four steps.
First, we will determine a spacetime region $u \subset \mathcal{U}$ where typical strong, local gravity solutions can be used consistently in a cosmological setting.
Second, we consider a {Kerr} BH, which requires extending the results of \S\ref{sec:zero-order} to additional boundaries.
Third, we consider a Schwarzschild constant-density sphere because this solution dominates many solutions of physical interest.
Finally, we consider a static de-Sitter sphere, because generalizations of such spheres have been proposed as BH replacements. 

\subsection{Spacetime Domain for Consistent Use of Strong, Local Gravity Approximations in Cosmological Settings}
\label{sec:birkhoff_oops}
First, fix some time of interest $\eta_0$ and consider some later time $\eta_0 + \Delta\eta$.
We may express Equation~(\ref{eqn:metric_picture_defn}) at $\eta_0 + \Delta\eta$ as
\begin{align}
  g_{\mu\nu} &= a^2(\eta_0 + \Delta\eta)\left[\eta_{\mu\nu} + \dots\right] \\
  &= \left\{a^2(\eta_0) + 2\Delta\eta\left(a \frac{\d a}{\d \eta}\right)\Bigg|_{\eta_0} + \dots\right\}\left[\eta_{\mu\nu} + \dots\right] \\
  &= a^2(\eta_0)\left\{1 + 2\Delta\eta\left(\frac{1}{a}\frac{\d a}{\d \eta}\right)\Bigg|_{\eta_0} + \dots\right\}\left[\eta_{\mu\nu} + \dots\right].
\end{align}
If we define a reciprocal timescale
\begin{align}
  H_\eta^0 \equiv \left(\frac{1}{a}\frac{\d a}{\d \eta}\right)\Bigg|_{\eta_0}, 
\end{align}
then the series expansion of Equation~(\ref{eqn:metric_picture_defn}) about $\eta_0$ gives
\begin{align}
  g_{\mu\nu}\Big|_u = a^2(\eta_0)\left[\eta_{\mu\nu} + \lim_{N\to\infty}\sum_{n=1}^N \epsilon^n h^{(n)}_{\mu\nu}(\eta, \vec{x}) + O\left(\Delta\eta H^0_\eta\right) \right]. \label{eqn:flattened_scaled}
\end{align}
Now choose $\Delta \eta$ such that the correction is small
\begin{align}
  \Delta \eta &\ll \left(H^0_\eta\right)^{-1}.   \label{eqn:short_eta_interval}
\end{align}
For any fixed $\epsilon < 1$, the $O\left(\Delta\eta H^0_\eta\right)$ term allows the approximation to be cut off at finite $N$
\begin{align}
  N < \frac{\log\left[\Delta\eta H^0_\eta\right]}{\log\epsilon}. \label{eqn:Delta_eta_constraint}
\end{align}
For physical reasons, we may choose a spatial volume $v \subset \mathcal{V}$, whose light-crossing time is less than $\Delta \eta$
\begin{align}
  v^{1/3} < \Delta \eta.
\end{align}
Note that mathematically, however, the following arguments will not depend on this spatial restriction.
We will restrict our attention to $\eta \in [\eta_0, \eta_0 + \Delta \eta]$ and define a spacetime domain $u \subset \mathcal{U}$ 
\begin{align}
  u \equiv [\eta_0, \eta_0 + \Delta\eta] \times v. \label{eqn:little_u}
\end{align}
Note that $u$ is a spacetime ``hockey puck,'' which encloses a region of the spatial slice $\mathcal{V}$ required for the spatial averages in Equation~(\ref{eqn:friedmanns_equation}).
Under a constant rescaling of the representation
\begin{align}
  \eta' \equiv \eta a(\eta_0) \qquad \vec{x}' \equiv \vec{x}a(\eta_0), \label{eqn:unit_rescaling}
\end{align}
Equation~(\ref{eqn:flattened_scaled}) becomes
\begin{align}
 g_{\mu\nu}\Big|_u = \eta_{\mu\nu} + \lim_{N \to \infty} \sum_{n=1}^N \epsilon^n h^{(n)}_{\mu\nu}\left(\eta', \vec{x}'\right) + O\left(\Delta \eta H^0_\eta\right).
 \label{eqn:flattened}  
\end{align}
This is the same metric, so the dynamics are unchanged.
In other words, it may be possible to use asymptotically flat models within a universe described by Equation~(\ref{eqn:metric_picture_defn}).
Conclusions drawn from such models, however, are only valid on the restricted spacetime $u$.
In particular, conclusions drawn on $u$ cannot be trivially extended off of $u$.
We summarize for clarity.
If the real world satisfies Assumptions 1 and 2, then for short intervals asymptotically flat models are permissible, but only for short intervals.

{ 
This is another reason why Birkhoff's theorem cannot be used to infer cosmological behavior.}
Given a sequence of times $\{\eta_j\}$ and suitable symmetry, one can use Birkhoff's theorem to construct a sequence of Schwarzschild exterior spacetimes characterized uniquely by a sequence of masses $\{M_j\}$.
Without global knowledge of the actual metric, however, no relation between the $M_j$ can be established.

We emphasize, however, that the converse relation holds.
In other words, it is always correct to proceed globally from Equation~(\ref{eqn:metric_picture_defn}) to locally, as represented by Equation~(\ref{eqn:flattened}).
This follows because $u \subset \mathcal{U}$.
This means Friedmann's equation remains valid at all events in $u$, where the metric takes the form given in Equation~(\ref{eqn:flattened}).
This means that any source on $u$, which produces a metric of the form Equation~(\ref{eqn:flattened}), will influence the dynamics of $a(\eta)$ at $\eta_0$, in the manner given by Equation~(\ref{eqn:friedmanns_equation}).

For all subsequent discussion, {we} fix some time $\eta_0$ and {we fix some} $\epsilon < 1$.
Because we have the large-scale solution by virtue of Assumptions 1 and 2, we have a well-defined $H^0_\eta(\eta_0)$.
Now choose $\Delta \eta$ and $N > 0$ so that Equation~(\ref{eqn:Delta_eta_constraint}) is satisfied.
We now have a well-defined $u$ given by Equation~(\ref{eqn:little_u}).
Strong sources defined in $u$ will the contribute to the dynamics of $a(\eta_0)$ through the spatial average over $u$ at $\eta_0$, as required by Friedmann's equations given in Equation~(\ref{eqn:friedmanns_equation}).
We now study some specific sources of interest.
\bigskip
\subsection{The Kerr Spacetime}
\label{sec:cutouts}
{ %\textbf{
LIGO \citep[e.g.][]{abbott2016observation} and the \citet{event2019first} have established the existence of ultracompact objects consistent with the Kerr geometry.
In this section, we will establish that the cosmological formalism developed continues to operate near to ultrarelativistic, spinning, sources.
Since the Kerr spacetime is asymptotically flat, the considerations of \S\ref{sec:birkhoff_oops} necessarily restrict use of the Kerr geometry to timescales $\ll 1/H_\eta^0$.
In particular, we determine an appropriate submanifold $u' \subset u$ defined in Equation~(\ref{eqn:little_u}), where the Kerr spacetime satisfies Assumptions 1 and 2 in the approximate sense of Equation~(\ref{eqn:flattened}).
For a convenient reference on the Kerr spacetime, we refer the reader to \citet{visser2007kerr}.

In Kerr-Schild coordinates, the Kerr metric is
\begin{align}
  \d s^2 = \left(\eta_{\mu\nu} + \frac{R_s r(\vec{x})^3}{r(\vec{x})^4 + A^2 z^2} \ell_\mu \ell_\nu\right) \d x^\mu \d x^\nu,
  \label{eqn:kerr}
\end{align}
where $R_s$ is the Schwarzschild radius for some fixed mass $M$, $A$ encodes dimensionful information about the spin,\footnote{\citet{visser2007kerr} uses $a$, but because we have used $a$ for the RW scale factor, we have switched to $A$} and $r(\vec{x})$ is a function of the position coordinates $x$, $y$, and $z$.
We will henceforth drop the explicit indication of $\vec{x}$ dependence for $r$.
The covector $\ell_\mu$ is defined to be null with respect to $\eta_{\mu\nu}$ and has the form
\begin{align}
  \ell_\mu &\equiv \left(1, \frac{rx + Ay}{r^2 + A^2}, \frac{ry - Ax}{r^2 + A^2}, \frac{z}{r}\right).
\end{align}
Since $\eta^{\mu\nu}\ell_\mu \ell_\nu \equiv 0$, it follows that
\begin{align}
  1 = \left(\frac{rx + Ay}{r^2 + A^2}\right)^2 + \left(\frac{ry - Ax}{r^2 + A^2}\right)^2 + \left(\frac{z}{r}\right)^2,
\end{align}
which implicitly defines $r$.
Since each term on the right hand side is positive, we see that the components of $\ell_\mu$ satisfy
\begin{align}
  |\ell_\mu| \leqslant 1.
\end{align}
It follows immediately that
\begin{align}
  \ell_k\ell_j \leqslant \ell_0|\ell_j| \leqslant \ell_0\ell_0 = 1. \label{eqn:ell_inequality}
\end{align}
To avoid a naked singularity, we require that
\begin{align}
  A < M,
\end{align}
which maintains real-valued solutions for the horizon surfaces.

By Equation~(\ref{eqn:ell_inequality}), a necessary and sufficient condition to write the metric in the form of Equation~(\ref{eqn:flattened}) is
\begin{align}
    \frac{R_s r^3}{r^4 + A^2 z^2} \leqslant \epsilon.
\end{align}
This restriction is saturated when 
\begin{align}
  \epsilon r^4 - R_s r^3 + \epsilon A^2 z^2 = 0.
\end{align}
By Descartes' rule of signs, this equation implicitly defines two (or zero) non-intersecting surfaces $r_\pm(\vec{x})$.
As $\epsilon \to 1$, $r_+$ and $r_-$ become, respectively, the outer and inner ergosurfaces of Kerr.
Define a submanifold $u' \subset u$ such that
\begin{align}
  u' \equiv [\eta_0, \eta_0 + \Delta \eta] \times v~/~\left\{ \vec{x} : \epsilon\left(r_+^4 + A^2 z^2\right) - R_s r_+^3 < 0 \right\}. \label{eqn:kerr_domain}
\end{align}
Then $u'$ has an additional boundary, which becomes the outer ergosurface as $\epsilon \to 1$.
We now establish that this additional boundary will not alter Friedmann's equations.
As discussed in Appendix~\ref{sec:boundary_term}, the dynamically relevant contribution to Friedmann's acceleration equation from this boundary is the second term in Equation~(\ref{eqn:boundary-split}).
This term continues to vanish at zero order on $u'$ because of the RW model symmetries.
As shown in Equation~(\ref{eqn:conservation_boundary}), an additional term also appears in the conservation equation
\begin{align}
  \int \epsilon^0(\eta)a^4\Bigg\{\int_{\partial u'} \tensor{T}{^k_0}(\eta, \vec{x})~\d^2x_k\Bigg\}~\d\eta.
\end{align}
Since Kerr is a vacuum solution, however, $\tensor{T}{^k_0} = 0$ on $\partial u'$ and this term also vanishes.

Now that we have a suitable domain, we may write Equation~(\ref{eqn:kerr}) in the form of Equation~(\ref{eqn:flattened}):
\begin{align}
  h_{\mu\nu}^{(1)} &\equiv \left(\frac{R_s}{r}\right) \left[1 + \frac{A^2z^2}{r^4}\right]^{-1}\ell_\mu\ell_\nu\\
  h_{\mu\nu}^{(n > 1)} &\equiv 0.
\end{align}
In summary, we have established a domain $u'$ where the Kerr solution can be accommodated under Assumptions 1 and 2.
This implies that the spatial averages appearing in Friedmann's equation remain valid in the ultrarelativistic vicinity of a Kerr BH. 
The domain $u'$ can be taken to include all but the region enclosed by the outer ergosurface.
Consequently, for the domain we have constructed, a Kerr BH contributes \emph{nothing} to Friedmann's equations.
Local observers in $u'$ will still perceive a Kerr BH with mass $M$ and spin $A$ for $r_+ < r < v^{1/3}$ and $\eta \in [\eta_0, \eta_0 + \Delta \eta]$.
This is unsurprising, because the cutting procedure we have performed is mathematical.
\citet{visser2007kerr} emphasizes that the inner horizon and the enclosed inner ergoregion are extremely pathological and should not be regarded as physically relevant.
Our result suggests that, additionally, the entire region below the outer ergosurface should be replaced with a distinct interior solution.
}

\subsection{Cosmological Contribution of Interior Solutions: Typical Astrophysical Sources}
\label{sec:schwarzschild_interior}
In this section, we show that the interior region of Schwarzschild's constant-density sphere satisfies Assumptions 1 and 2.
The exterior region of Schwarzschild's constant-density sphere is {an $A = 0$ Kerr solution (i.e. Schwarzschild's BH)}, which has already been treated.
The constant-density solution can be chosen to dominate the actual energy densities for many spherically symmetric, static, sources.  
We will show that, for such objects with physical radius $R > 3GM$, all energy densities and pressures in $u$ will influence $a(\eta)$ according to Equation~(\ref{eqn:friedmanns_equation}).
This will establish that localized pressures, interior to a large class of compact and relativistic objects, contribute to the global Friedmann average.

The Schwarzschild constant-density sphere on $B_3(0, R)$ may be expressed as
\begin{align}
  \d s^2 = -\exp(2\Phi)~\d \eta^2 + \exp(2\Lambda)~\d r^2 + r^2\d\Omega^2 \label{eqn:use_the_schwarz}.
\end{align}
The functions $\Phi$ and $\Lambda$ are defined in terms of the Schwarzschild radius $R_s$ and physical radius $R$ of the object
\begin{align}
  \exp\left(2\Phi\right) &\equiv \frac{1}{4}\left[3\sqrt{1 - \frac{R_s}{R}} - \sqrt{1 - \frac{R_s}{R}\left(\frac{r}{R}\right)^2}\right]^2 \label{eqn:exp2Phi}\\
  \exp\left(2\Lambda\right) &\equiv \left[1 - \left(\frac{r}{R}\right)^2\frac{R_s}{R}\right]^{-1}.
\end{align}
We will show that there exists a natural $\epsilon$ such that the coefficients of a Taylor expansion of Equation~(\ref{eqn:use_the_schwarz}) take the form of Equation~(\ref{eqn:flattened}).

Denote the Taylor expansion coefficients of $\sqrt{1-x}$ on $0 \leqslant x < 1$ by $q_j$
\begin{align}
  \sqrt{1-x} &= 1 - \frac{x}{2} - \sum_{j=2}^\infty \frac{(2j-3)!}{2^{2(j-1)}j!(j-2)!}x^j \label{eqn:sqrt_expansion}\\
  &\equiv q_0 - q_1 x -  \sum_{j=2}^\infty q_jx^j.
\end{align}
Furthermore, extend the notation to negative indices by
\begin{align}
  q_k \equiv 0 \qquad k < 0.
\end{align}
Then it can be shown that the expansion of Equation~(\ref{eqn:exp2Phi}) in $R_s/R$ becomes
\begin{align}
  \begin{split}
    \exp(2\Phi) = 1 -& \frac{R_s}{R}\left[\frac{3}{2} - \frac{1}{2}\left(\frac{r}{R}\right)^2 \right] \\
    &- \frac{3}{2}\sum_{n=2}^\infty \left(\frac{R_s}{R}\right)^n\left[\sum_{j=0}^\infty q_j q_{n-j}\left(\frac{r}{R}\right)^{2j}\right].
    \end{split}
\end{align}
The terms in these infinite sums can be commuted and rearranged because $\sqrt{1-x}$ converges uniformly for $|x| < 1$.
To guarantee that the coefficient of the linear term remains bounded by 1 on $r \in [0,R]$, we must have
\begin{align}
  \epsilon \geqslant \frac{3R_s}{2R}.  \label{eqn:defn_incompressible_epsilon}
\end{align}
Since $\epsilon < 1$, this becomes a restriction on the object's physical radius
\begin{align}
  R > 3GM.
\end{align}
Note that this is the radius of the innermost stable orbit for a photon.\footnote{Compare this with the requirement $R > 9GM/4$ for the existence of static solutions, given by \citet[][\S6.2]{WaldGR} as the Buchdahl bound.}
Saturating this bound, we find
\begin{align}
  \begin{split}
    \exp(2\Phi) = 1 -& \epsilon\left[1 - \frac{1}{3}\left(\frac{r}{R}\right)^2 \right] \\
    &- \sum_{n=2}^\infty \epsilon^n\left[\sum_{j=0}^\infty q_j q_{n-j}\left(\frac{2}{3}\right)^{n-1}\left(\frac{r}{R}\right)^{2j}\right].
    \end{split}
\end{align}
To bound the remaining coefficients of $\epsilon^n$, note that for each coefficient in $n$,
\begin{align}
  \sum_{j=0}^\infty q_j q_{n-j}\left(\frac{2}{3}\right)^{n-1}\left(\frac{r}{R}\right)^{2j} &< \sum_{j=0}^\infty q_j\left(\frac{r}{R}\right)^{2j} = \sqrt{1 - \left(\frac{r}{R}\right)^2}.
\end{align}
The inequality follows because $q_{n-j} < 1$ for all $j$.
Since $\sqrt{1-x} \leqslant 1$ for $x\in[0,1]$, each coefficient of $\epsilon^n$ for $n \geqslant 2$ is bounded above by 1 on $r\in[0,R]$.
This was to be shown.
The result for $\exp(2\Lambda)$ follows immediately from use of the binomial expansion, followed by substitution of Equation~(\ref{eqn:defn_incompressible_epsilon}).
In Appendix~\ref{sec:constant_density_einstein}, we establish that the Schwarzschild constant-density sphere also satisfies Assumption 2.

In summary, we have established that the Schwarzschild constant-density sphere is of the form Equation~(\ref{eqn:flattened}).
In other words, all pressures interior to any astrophysical object bounded by the Schwarzschild constant-density sphere solution will contribute to the cosmological average in Equation~(\ref{eqn:friedmanns_equation}).

\subsection{Cosmological Contribution of Interior Solutions: Static de-Sitter Sphere}
\label{sec:desitter}
In this section, we show that the interior of an isolated de-Sitter sphere satisfies Assumptions 1 and 2.
The exterior region of such a sphere is {an $A = 0$ Kerr solution (i.e. Schwarzschild's BH)}, which has already been treated.
The isolated de-Sitter sphere is the simplest model of a GEneric Object of Dark Energy (GEODE).
Such objects, like the solution of \citet{dymnikova1992vacuum} or the gravastar of \citet{mazur2015surface}, have been proposed as possible BH replacements.
Related, but dynamic, GEODEs called ``vacuum bubbles,'' have also been considered as inflationary relics by \citet{berezin1987dynamics}.
We will demonstrate shortly that the strong negative pressure inside the de-Sitter sphere will influence $a(\eta)$ according to Equation~(\ref{eqn:friedmanns_equation}).
This will establish that a physically realistic GEODE could contribute to the cosmological $\mathcal{P} = -\rho$.

Consider a de-Sitter patch in static coordinates 
\begin{align}
  \d s^2 = -\left(1 - \frac{r'^2}{R_s^2}\right)\d\eta'^2 + \left(1 - \frac{r'^2}{R_s^2}\right)^{-1}\d r'^2 + r'^2\d\Omega^2.
  \label{eqn:desitter}
\end{align}
We have introduced primes on the $\eta'$ and $r'$ coordinates for reasons that will become apparent.
Denote these coordinates by $\Xi'$.
This patch admits a timelike Killing vector field for $r' < R_s$.
This means that {a sphere of Dark Energy with radius $R < R_s$} is static, regardless of the coordinate representation.

Again, expand in a series, then multiply and divide by $\epsilon$:
\begin{align}
   g_{\mu\nu} = \eta_{\mu\nu}(\eta', \vec{x}') + \epsilon^2 \left(\frac{r'}{R_s\epsilon}\right)^2\tensor{\delta}{^0_\mu}\tensor{\delta}{^0_\nu} + \sum_{n=1}^\infty \epsilon^{2n} \left(\frac{r'}{R_s\epsilon}\right)^{2n}\tensor{\delta}{^1_\mu}\tensor{\delta}{^1_\nu}.
  \label{eqn:desitter_rep}
\end{align}
We see that $\left|h_{\mu\nu}^{(n)}\right| \leqslant 1$, and so Equation~(\ref{eqn:desitter_rep}) satisfies Equation~(\ref{eqn:flattened}), provided that
\begin{align}
  r' \leqslant R_s\epsilon.
\end{align}
In other words, Equation~(\ref{eqn:desitter}) has a convergent representation on $B_3(0, R_s\epsilon)$.
Contrast this situation with the Schwarzschild BH, which {converges} on $v~/~B_3(0, R_s\epsilon^{-1})$.
Because Einstein's equations hold everywhere on $\mathcal{U}$, we know there exist coordinates $\zeta$, which cover $r' > R_s\epsilon$.
This means there exists $z > 0$ such that 
\begin{align}
  \mathscr{O}' &\equiv \mathrm{dom}(\Xi') \cap \mathrm{dom}(\zeta) \\
  &= \left(\eta_0', \eta_0' + \Delta \eta'\right) \times \left(R_s\epsilon - z, R_s\epsilon\right) \times S^2
\end{align}
is the overlap between the $\Xi'$ and $\zeta$ charts.
We will continue to use the $\Xi'$ chart on this region.

The relation between the $\Xi$ coordinates of {the exterior Schwarzschild spacetime} and the $\Xi'$ coordinates of the de-Sitter sphere is determined by \citet[][Eqns.~(5.1--5.3)]{mazur2015surface}.
They find that $\Xi$ and $\zeta$ are the same, and that $\Xi'$ satisfies
\begin{align}
  r' &= r \\
  \d \eta' &= 2~\d\eta. \label{eqn:interior_clock}
\end{align}
In other words, the spatial slices are unaltered, but time interior to the sphere runs twice as fast.
This means that our static approximation is only valid for $\Delta \eta / 2$.
Apart from a reduction in $v$ to $v/8$, nothing changes with respect to the spatial-slice integration.
We may integrate Equation~(\ref{eqn:interior_clock}) to find
\begin{align}
  \eta'(\eta) &= 2\eta + C,
\end{align}
where $C$ is an integration constant.
Note that we may choose $C \equiv -\eta_0$ so that
\begin{align}
  a\left[\eta'(\eta_0)\right] = a(\eta_0).
\end{align}
This shows that the unit redefinition used to write Equation~(\ref{eqn:flattened}) remains consistent inside the sphere.

We now cut out the $r > R_s\epsilon - z/2$ region, leaving a boundary at the cut radius.
It again follows that Equation~(\ref{eqn:boundary-split}) continues to vanish at zero order.
The required conservation condition in Equation~(\ref{eqn:conservation-volume}) continues to hold because $\tensor{T}{^\mu_\nu} \propto \tensor{\delta}{^\mu_\nu}$, and so the contribution vanishes identically. 
Consider the domain $u'' \subset u$
\begin{align}
  u'' \equiv \left[\eta_0, \eta_0 + \frac{\Delta\eta}{2}\right] \times \frac{v}{8}~/~\left\{\left[R_s\epsilon - \frac{z}{2}, \frac{R_s}{\epsilon} + \frac{z}{2}\right] \times S^2\right\}. \label{eqn:desitter_domain}
\end{align}
We have established that Equation~(\ref{eqn:flattened}) is satisfied everywhere on $u''$ and that there are no additional boundary contributions to Equation~(\ref{eqn:friedmanns_equation}).
Finally, static de-Sitter space satisfies Assumption 2 because its stress tensor is constant.
By adjusting $z$ and $\epsilon$ in {Equation~(\ref{eqn:desitter_domain})}, we conclude that nearly all of the de-Sitter region contributes $\mathcal{P}=-\rho$ to Equation~(\ref{eqn:friedmanns_equation}).

{
\section{Cosmological energy shifts}
\label{sec:consequences}
Friedmann's equation in \S\ref{sec:zero-order} clarifies that local pressure contributions, which do not vanish upon spatial averaging, affect the scale factor $a(t)$.
In \S\ref{sec:strong}, we established that the framework of \S\ref{sec:zero-order} remains valid very near, and often within, ultrarelativistic sources.
In this section, we show how any source that contributes to the cosmologically averaged pressure must itself evolve cosmologically.

In the following, we will use the word ``object'' to refer to bound systems like stars and clusters.
How a particular object responds to the scale factor is entirely dependent on the object.
To see this, consider a population of generic objects.
Let $\mathcal{P}(\eta)$ denote the spatially averaged pressure over the population.
Let $\rho(\eta)$ denote the spatially averaged energy density over the population.
Define the equation of state of the contribution to be
\begin{align}
  w(\eta) \equiv \frac{\mathcal{P(\eta)}}{\rho(\eta)}. \label{eqn:w_defn}
\end{align}
Conservation of stress-energy, given in Equation~(\ref{eqn:zero-order-conservation}), describes the temporal evolution of these averaged quantities
\begin{align}
  -\frac{\d\rho}{\d \eta} - \frac{1}{a}\frac{\d a}{\d \eta}\left(3\rho + 3\mathcal{P}\right) = 0.
\end{align}
From the definition of $w(\eta)$, we may write
\begin{align}
    -\frac{\d\rho}{\d \eta} - \frac{3\rho}{a}\frac{\d a}{\d \eta}(w + 1) = 0.
\end{align}
Switching to the scale factor $a$ as the independent variable gives the separable differential equation
\begin{align}
  \frac{\d \rho}{\rho} = -\frac{3[w(a) + 1]~\d a}{a}. \label{eqn:cons_nonconstant}
\end{align}
Now suppose that $w$ is constant.
Then we may integrate Equation~(\ref{eqn:cons_nonconstant}) to find the standard result,
\begin{align}
  \rho(a) \propto a^{-3[w + 1]}. \label{eqn:nut}
\end{align}

For simplicity, suppose all of the objects have the same comoving energy $E$.
The energy density can then be written in terms of the physical number density $\mathcal{N}$ of the object population
\begin{align}
  \rho(a) = E \mathcal{N}.
\end{align}
The number density dilutes with the expansion
\begin{align}
  \mathcal{N} \propto \frac{1}{a^3} \label{eqn:dNdV},
\end{align}
because objects either comove or belong to structures that comove.
Combining Equation~(\ref{eqn:nut}) and Equation~(\ref{eqn:dNdV}) gives
\begin{align}
  E \propto a^{-3w} \label{eqn:hardnut}.
\end{align}
This result is again expected.
For example, if the objects are photons, $w=1/3$ and so $E \propto 1/a$, which is the photon redshift.

Our result, however, applies to all objects.
This consequence follows from Assumptions 1 and 2, and the EH action.
When $w(a)\neq 0$, the averaged quantities $\rho(\eta)$ and $\mathcal{P}(\eta)$ evolve cosmologically.
Thus, any material that contributes non-vanishing pressure to Equation~(\ref{eqn:friedmanns_equation}) and Equation~(\ref{eqn:zero-order-conservation}) must also evolve cosmologically.
The evolution of $E$ persists when $w(a)$ varies in time, as can be seen from Equation~(\ref{eqn:cons_nonconstant}).
For compact objects, the particular value of $w$ is strongly dependent on the non-gravitational fields $\mathbf{\Psi}$ that define the material.
The cosmological energy shift is completely unaffected by the spatial distribution of material in the universe.

\subsection{Observations}
The cosmological evolution of local energies is a necessary consequence of Assumptions 1 and 2, and the EH action.
The essential question is now observational: can this effect be measured?
Because this section is observational, it is sometimes clearer to use redshift $z$ as a time variable.
We thus convert Equation~(\ref{eqn:hardnut}) to redshift
\begin{align}
  E_i \propto (1+z)^{3w}, 
\end{align}
using that $a = (1+z)^{-1}$.
Where appropriate, we use Planck collaboration cosmological parameters from \citet{aghanim2018planck}: $H_0 = 67.4 \pm 0.5~\mathrm{km}~\mathrm{Mpc}^{-1}~\mathrm{s}^{-1}$, $\Omega_b = (2.24 \pm 0.01)\times 10^{-2} h^{-2} \simeq 0.05$, $\Omega_\Lambda \equiv 1 - \Omega_m = 0.685 \pm 0.007$.
We now consider specific model objects relevant for astrophysical observation, using fixed $w$ approximations.
We proceed in order of increasingly relativistic $w$.

\subsubsection{Stars}
Consider a population of typical stars, at fixed (comoving) coordinate positions.
Each typical star will contribute an extremely small positive pressure to the cosmological average.
This follows because the pressure is everywhere positive within a star, and so an integral over the stellar pressure cannot vanish.
This is true even in simplified stellar models, where fluid packets are radially static.
Each packet's contribution to the pressure is non-zero because gravitational momentum flux is not included in $T_{\mu\nu}$.
The equation of state $w$ for a typical star can be approximated from the ideal gas law,
\begin{align}
  w_\mathrm{star} \sim \frac{k T}{m_\mathrm{p}}
\end{align}
where $k$ is Boltzmann's constant, $T$ is the temperature of the star, and $m_\mathrm{p}$ is the proton mass.
To develop an upper bound, set $T \sim 10^6~\mathrm{K}$, representative of core temperatures.
Then
\begin{align}
  w_\mathrm{star} \sim 10^{-7}.
\end{align}
Thus, the energy (as perceived by RW observers) of Sun-like stars cosmologically evolves as
\begin{align}
  E_\mathrm{star} \propto (1+z)^{3\times 10^{-7}}.
\end{align}
How large is this effect?
Suppose a star is produced at $z_i=2$ and observed at $z_f=0$.
Define the fractional shift in energy as
\begin{align}
  \frac{\Delta E}{E} \equiv \frac{E(z_f) - E(z_i)}{E(z_i)}, 
\end{align}
then
\begin{align}
  \frac{\Delta E_\mathrm{star}}{E_\mathrm{star}} &= \left(\frac{1}{1 + 2}\right)^{3\times 10^{-7}} - 1 \simeq -10^{-7}.
\end{align}
This change occurs over $10~\mathrm{Gyr}$ from $z_i=2$ to $z_f=0$.
In other words, it is dominated by other stellar processes and is thus unobservable.
This also establishes that the effect is unobservable for any other material with $w < w_\mathrm{star}$.

What is the reciprocal effect on the zero-order expansion?
Stars contribute $\sim 2\Omega_b/5$ to Friedmann's equation.
The cumulative adjustment to $\rho(a)$ from a conservative first light of $z_i=40$ to $z_f=0$ is then $\sim -10^{-8}$.
The effect is unobservable.

\subsubsection{Galaxy Clusters}
Consider a population of clusters, at fixed (comoving) coordinate positions.
The velocity dispersion of clusters is roughly bounded by $\sigma \sim 10^3~\mathrm{km}~\mathrm{s}^{-1}$.
This implies a $w_\mathrm{cluster} \sim 10^{-5}$ by ideal gas arguments.
Thus, the energy (as perceived by RW observers) of clusters cosmologically evolves as
\begin{align}
  E_\mathrm{cluster} \propto (1+z)^{3\times 10^{-5}}.
\end{align}
How large is this effect?
A typical length scale for a cluster is $\sim 1~\mathrm{Mpc}$, giving a light-crossing time of $\sim 3~\mathrm{Myr}$.
In redshift, this could be between $z_i = 1.001$ and $z_f = 1$.
During this time, the fractional shift in energy of the cluster is
\begin{align}
  \frac{\Delta E_\mathrm{cluster}}{E_\mathrm{cluster}} &= \left(\frac{1 + 1}{1 + 1.001}\right)^{3\times 10^{-5}} - 1 \simeq -10^{-8}.
\end{align}
For comparison, during a photon transit completing at $z_f=2$, the cluster energy shifts less than $10^{-8}$.
In other words, the correction to any integrated Sachs-Wolfe effect through clusters is unobservable.
Similarly, corrections to gravitational lensing are unobservable.

What is the reciprocal effect on the zero-order expansion?
The energy density in galaxy clusters is certainly bounded above by $\Omega_b$. 
The cumulative adjustment to $\rho(a)$ from a conservative first cluster of $z_i=4$ to $z_f=0$ is then $\sim -10^{-6}$.
The effect is unobservable at present and upcoming sensitivities.

\subsubsection{GEneric Objects of Dark Energy (GEODEs)}
As discussed in \S\ref{sec:desitter}, GEODEs are explicit GR solutions, which schematically resemble the static de-Sitter sphere.
Before gravitational-wave observations of ultrarelativistic object mergers, GEODEs were of theoretical interest because they are often free of physical singularities and horizons.
In other words, they are regular solutions for gravitational-collapse remnants, which resolve the BH Information Paradox.

Consider a population of GEODEs positioned at fixed (comoving) coordinate positions.
Recall that the equation of state of a de-Sitter sphere is
\begin{align}
  w_\mathrm{dS} = -1.
\end{align}
This equation of state is appropriate if the GEODE edge does not contribute significantly to the population-averaged $\mathcal{P}$.
GEODE material is maximally relativistic: it saturates the dominant energy condition.
The material is also under tension, instead of under compression: the sign of $w$ is inverted from all previously considered cases.
This leads to very useful consequences.

The energy (as perceived by RW observers) of the GEODEs cosmologically evolves as
\begin{align}
  E_\mathrm{dS} \propto (1+z)^{-3}.
\end{align}
Unlike the cases previously considered, this shift is significant and \emph{acts to amplify the energy.}
In other words, GEODEs cosmologically blueshift.
The effect is analogous to the photon redshift.

How large is this effect?
Consider a $3\mathrm{M}_\odot$ GEODE formed from stellar gravitational collapse at $z=1.5$, near the peak in comoving star formation rate density \citep[e.g.][]{MadauDickinson2014}.
Its mass, observed in a binary merger at $z=0.1$, will be
\begin{align}
  E_\mathrm{dS} = 3\mathrm{M}_\odot \left(\frac{1 + 1.5}{1 + 0.1}\right)^3 = 35.2\mathrm{M}_\odot.
\end{align}
This compares favorably with the masses observed by LIGO.
A thorough exploration of the observational signatures of a remnant GEODE population, with respect to gravitational-wave observatories, is given by \citet{cnf2019part3}.
They find that the scenario is consistent with the LIGO GWTC-1 observed mass function.
The cosmological blueshift of a GEODE can achieve the required masses with the standard common envelope binary formation channel.
It does not require low-metallicity regions or prohibitively large progenitor stars.

What is the reciprocal effect on the zero-order expansion?
Suppose $1\%$ of all stellar material collapsed to Population III GEODEs, instead of traditional black holes,
\begin{align}
  \Omega_\mathrm{collapse} \sim 2\times 10^{-4}.
\end{align}
For simplicity, assume that this happens at $z \sim 14$: after the Dark Ages but before reionization.
The shift in energy of this population, observed at the present day, gives rise to
\begin{align}
  \Omega_\mathrm{GEODE} &= \Omega_\mathrm{collapse} (1 + 14)^3 \\
  &= 0.675.
\end{align}
Note that $\rho_\mathrm{GEODE}$ is essentially constant within Friedmann's equations.
This follows because $w_\mathrm{dS} = -1$, so that Equation~(\ref{eqn:nut}) gives no time evolution of the physical density.
In other words, the GEODEs dilute in number $\propto a^{-3}$ while gaining in mass $\propto a^3$.
The result is an apparent cosmological constant, which compares favorably with the observed $\Omega_\Lambda$.

A first exploration of the observational signatures of Population III GEODEs, with respect to cosmological observables, is given by \citet{CrokerThesis2018}.
They find that such a scenario is flexible enough to reproduce the observed late-time accelerated expansion, resolving the coincidence problem.

\subsection{Discussion}
We have shown that non-relativistic material exhibits no observable shift.
For relativistic objects, however, the energy shift can lead to measurable consequences.
Neutron star (NS) material is highly relativistic \citep[e.g.][]{abbott2018gw170817}, with an object-averaged $0.05 \lesssim w_\mathrm{NS} \lesssim 0.1$.
Given that binary pulsar dynamics are often measured to exquisite precision,
it may be possible to measure the cosmological shift in pulsar systems.

Since the cosmological shift for positive $w_\mathrm{NS}$ appears as an energy loss, one can use a result of \citet[][Equation~(4.1)]{damour1991orbital} to estimate the shift in orbital period decay 
\begin{align}
  \Delta\dot{P}_\mathrm{b,cos} &= +6w_\mathrm{NS}H P_\mathrm{b}. \label{eqn:spurious_shift}
\end{align}
Using $P_\mathrm{b}$ as reported by \citet[][Table~3]{weisberg2010timing} for the Hulse-Taylor binary pulsar, we find
\begin{align}
  \Delta\dot{P}_\mathrm{b,cos} = (2.7 \pm 0.9) \times 10^{-14},
\end{align}
at the present day.
How does this shift compare to current measurements of the Hulse-Taylor system?
From \citet[][Equation~(5) and Table~3]{weisberg2010timing}, the kinematically corrected orbital period decay is
\begin{align}
  \dot{P}_\mathrm{b} - \Delta \dot{P}_\mathrm{b,gal} = (-2.396 \pm 0.005) \times 10^{-12}. \label{eqn:ht_decay}
\end{align}
The shift in orbital period decay at the present, due to cosmological evolution of NS energies, would appear to be $\sim 1\%$.

The central value of Equation~(\ref{eqn:ht_decay}) agrees with the GR radiative loss prediction to $0.2\%$.  
So a cosmological effect of the estimated magnitude has not been observed.
Since \citet{damour1991orbital} assumed a flat spacetime, however, it follows from Equation~(\ref{eqn:flattened}) that the metric is known only to $O(H P_\mathrm{b})$.
In other words, the estimated shift given by Equation~(\ref{eqn:spurious_shift}) is dominated by the error introduced by working under an asymptotically flat approximation within an RW cosmology.
An investigation of binary pulsar systems with cosmologically evolving mass is the topic of future work.

The contribution to the cosmologically averaged quantities, and the necessary local evolution, is highly dependent on the object model.
An instructive example is given by \citet{dymnikova1992vacuum}.
Dymnikova's object contains a very nearly Dark Energy interior, but a spatial average over her object gives exactly $\mathcal{P}=0$.
In her model, the ``skin'' acts like a vacuum vessel: positive pressures in the skin maintain the static character of the object.
A population of Dymnikova's objects would behave as cosmological dust, and exhibit no cosmological shift.

At present, no known GEODE solution strongly rotates.
As pointed out by J.~D.~Bjorken (2018, private discussion), vacuum cannot rotate because it has no privileged directions.
Given clear evidence of Kerr exterior spacetimes, all of the spin must reside in material outside of, but extremely near, the de-Sitter core.
GEODE solutions are often constructed to have a Schwarzschild exterior spacetime and thus cannot evolve cosmologically.
Given the considerations of \S\ref{sec:birkhoff_oops}, and the consequences in \S\ref{sec:consequences}, it seems essential to have adiabatically evolving object solutions that spatially asymptote to an arbitrary RW geometry.
The construction of a realistic GEODE model is an open question that is beyond the scope of this paper.}

\section{Conclusion}
\label{sec:conclusion}
In summary, we have derived the equations of motion for Friedmann cosmology, paying particular attention to the symmetry enforced by isotropy and homogeneity.
Working directly from the EH action maintains this symmetry, with respect to the source terms for Friedmann's equations.
Our main results, embodied in Equation~(\ref{eqn:friedmanns_equation}), follow from only two assumptions.
We assume that the metric and Einstein tensor have convergent series representations, in RW coordinates.
{
Contrary to some earlier literature, we find no influence on the source to Friedmann's equation from inhomogeneities in the spatial distribution of matter.
Contrary to other earlier literature, we find that every pressure source contributes to Friedmann's equations.}
{ Specifically}, the spatial average in Equation~(\ref{eqn:friedmanns_equation}) includes the {pressures interior to} compact objects.
{Evolution of the averaged quantities then implies a local energy shift of $a^{-3w}$ in any material that contributes to the averaged pressure.
  Here, $a$ is the RW scale factor and $w \equiv \mathcal{P}/\rho$ is the equation of state determined from the population-averaged pressure $\mathcal{P}$ and energy density $\rho$ of the objects.
  In non-relativistic material, the shift is too small to observe.
  The shift can be observable in relativistic material, if the equation of state exceeds $w \sim 0.01$.
  This result takes on particular significance when applied to Population III stellar-collapse remnants with $\mathcal{P} = -\rho$ interiors.
  A population of such stellar collapse remnants can shift in energy $\propto a^3$ while diluting in number density $\propto 1/a^3$.
  The population-averaged energy density is then effectively constant and readily produces the observed $\Omega_\Lambda$.}

\acknowledgments K.~S.~Croker warmly acknowledges:
N.~Kaiser~(ENS) for numerous conversations clarifying the historical understanding of pressure within FLRW models;
X.~Tata~(Hawai`i) for penetrating feedback and explicit revisions;
E.~Mottola (LANL) for traveling to Hawai`i for discussions concerning applications of this work and for pointing out omissions in early arguments;
{D.~Farrah (Hawai`i) for comments on clarity;}
J.~D.~Bjorken~(SLAC) for regular correspondence and encouragement; and
I.~Szapudi~(IfA) for criticism, which significantly strengthened the arguments presented.
Portions of this work were performed with financial support from the Fulbright U.S.
Student Program and the UH Vice Chancellor for Research.

\software{\href{http://maxima.sourceforge.net}{Maxima}~\citep{maxima}}

\begin{appendix}
  \section{Detailed discussion of the boundary term}
  \label{sec:boundary_term}
  By Stokes' theorem, we may write $\mathcal{B}$ as 
  \begin{align}
    \mathcal{B} &\propto \int_\mathcal{U} \partial_\mu \left\{\eta^{\mu\nu}\left[a \partial_\nu\delta a - \delta a \partial_\nu a \right]\right\}\d^4x \\
    &= \int_{\partial \mathcal{U}} \left\{\eta^{\mu\nu}\left[a \partial_\nu\delta a - \delta a \partial_\nu a \right]\right\}\d^3x_\mu, \label{eqn:bd_surface_integral}
  \end{align}
  where we ignore the factor $3/8\pi G$ for clarity.
  The contribution along the boundary of $\mathcal{U}$ can be broken down into two parts.
  An ``end-cap'' contribution from the future and past Cauchy surfaces (entire spatial 3-volumes at an instant of time) and a spatial 2-surface during the time interval considered
  \begin{align}
    \partial \mathcal{U} = \left[\mathcal{V} \times \{\eta_i, \eta_f\}\right] \cup \left[\partial \mathcal{V} \times [\eta_i, \eta_f] \right].
  \end{align}
  The intersection of these sets is $\{\eta_i, \eta_f\}\times\partial\mathcal{V}$, but it is two-dimensional and so it is a set of zero measure.
  In other words, the integral in Equation~(\ref{eqn:bd_surface_integral}) is three-dimensional and the integrand must be non-infinite, so this portion of the boundary contributes nothing.
  Thus, we may write 
  \begin{align}
    \begin{split}
      \mathcal{B} \propto -\int_{\mathcal{V}} \bigg(&a\partial_0\delta a - \delta a \partial_0 a\bigg)~\d^3x\Bigg|_{\eta_i}^{\eta_f} \\
      &+\int_{\eta_i}^{\eta_f} \int_{\partial \mathcal{V}} \eta^{k\nu}\bigg(a \partial_\nu \delta a - \delta a \partial_\nu a \bigg)~\d^2 x_k~\d \eta.
    \end{split} \label{eqn:boundary-split}
  \end{align}
  We have used Fubini's theorem to write the integrals as iterated integrals, permissible because $\mathcal{U}$ is bounded and $a$ is assumed to be well-behaved.
  The second term in this sum vanishes identically.
  This follows immediately because $a(\eta)$ and $\delta a(\eta)$ are functions of time alone (isotropy and homogeneity), so only temporal derivatives survive.
  This leaves only $\eta^{k0}$ terms, but these all vanish.
  We now expand the Cauchy surface ``end-cap'' contribution 
  \begin{align}
    \mathcal{B} &\propto \mathcal{V}\left(a\partial_0\delta a - \delta a \partial_0 a\right)\big|_{\eta_i}^{\eta_f}.
  \end{align}
  Physically, the usual procedure is to constrain the variations $\delta a(\eta)$ at the two endpoints.
  This leaves a single term 
  \begin{align}
    \mathcal{B} &\propto \mathcal{V}a\left(\partial_0\delta a\right) \big|_{\eta_i}^{\eta_f}.
    \label{eqn:residual_boundary}
  \end{align}
  Mathematically, if we take complete control over the variations $\delta a(\eta)$, we may always write them as mollified by a $C^\infty$ bump on some compact subset of $(\eta_i, \eta_f)$.
  This guarantees control of all derivatives at the endpoints, and can be used to destroy this term.
  If we do not assert this level of control, this final boundary term is just the York boundary term.

  To see this, we follow \citet[Equation
    (E.1.18)]{WaldGR} where the varied term of the EH action, which gives rise to the York term, is 
  \begin{align}
    \int_\mathcal{U} \nabla_\alpha\left(v_\beta g^{\alpha\beta}\right) \mathbf{\epsilon}.
  \end{align}
  For our RW ansatz, we find for \citet[Equation
    (E.1.16)]{WaldGR}
  \begin{align}
    v_\alpha &\equiv \nabla^\beta \delta g_{\alpha\beta} - g^{\gamma\delta}\nabla_\alpha\delta g_{\gamma\delta} = -6a^{-2}\df{(a\delta a)}{0}\tensor{\delta}{^0_\alpha} \\
    \mathbf{\epsilon} &\equiv \sqrt{-g}~\d^4 x = a^4~\d^4x.
  \end{align}
  By Stokes' theorem, we have 
  \begin{align}
    \int_\mathcal{U} \nabla_\alpha\left(v_\beta g^{\alpha\beta}\right) \mathbf{\epsilon} = \int_{\partial \mathcal{U}} v_\beta n_\alpha g^{\alpha\beta} \mathbf{e}
  \end{align}
  where, for our RW ansatz, we find 
  \begin{align}
    n_\alpha n_\beta g^{\alpha\beta} &\equiv -\tensor{\delta}{^0_\alpha} \tensor{\delta}{^0_\beta} \left\Vert n_\alpha \right\Vert ^2 a^{-2} \equiv -1 \implies \left\Vert n_\alpha\right\Vert = a \\
    n_\alpha v_\beta g^{\alpha\beta} &= 6a^{-4}\tensor{\delta}{^\alpha_0} \df{(a\delta a)}{0} n_\alpha \\
    \mathbf{e} &= \sqrt{h}~\d^3x = a^3~\d^3x.
  \end{align}
  Note that $h$ is Wald's notation for the induced metric on the hypersurface orthogonal to the timelike unit covector field $n_\alpha$.
  The resulting boundary term is 
  \begin{align}
    \int_{\partial \mathcal{U}} v_\beta n_\alpha g^{\alpha\beta} \mathbf{e} &= 6\mathcal{V}\df{(a \delta a)}{0}\big|_{\eta_i}^{\eta_f} \\
    &= 6\mathcal{V}\left(\df{a}{0}\delta a + a \df{\delta a}{0}\right)\big|_{\eta_i}^{\eta_f} \\
    &= 6\mathcal{V} a (\df{\delta a}{0})\big|_{\eta_i}^{\eta_f}, 
  \end{align}
  where the final equality follows because $\delta a$ vanishes at the temporal endpoints.
  This term is proportional to Equation~(\ref{eqn:residual_boundary}), which was to be shown.

  A term of this form remains because the gravitational Lagrangian, built from the Ricci scalar, contains second-order derivatives of the field variables.
  The established procedure, given by \citet[Equation
    (E.1.42)]{WaldGR}, is to just extend the EH action with the \emph{negative} of this term.
  Thus, this final boundary term vanishes by construction.
  The York term is often omitted from the action when working classically, as it is non-dynamical.

  \section{The Einstein Tensor of the Constant-Density Sphere}
  \label{sec:constant_density_einstein}
  In this section, we explicitly show that the Schwarzschild constant-density sphere of \S\ref{sec:schwarzschild_interior} satisfies Assumption 2 of \S\ref{sec:zero-order}.
  First, note that the constant-density sphere is static and spherically symmetric.
  This implies that $G_{\mu\nu} = 0$ when $\mu \neq \nu$.
  By Einstein's equations, it suffices to consider $T_{\mu\nu}$ to develop a bound on $G_{\mu\nu}$.
  First, we require a bound on the density of the sphere.
  From Equation~(\ref{eqn:defn_incompressible_epsilon}), the minimum physical radius $R$ must satisfy
  \begin{align}
    R > \frac{3R_s}{2} \label{eqn:minimum_incompressible_physical_radius}.
  \end{align}
  This implies a density
  \begin{align}
    \rho < \frac{1}{9\pi G R_s^2}.
  \end{align}
  The distance scale for variations in the gravitational field is determined by $R_s$.
  This gives
  \begin{align}
    G_{00} < \frac{8}{9}T_{00},
  \end{align}
  Because the density is constant, { it can be scaled so that} $G_{00}$ satisfies Assumption 2.
  
  We now develop the explicit bound on $G^{(n)}_{kk}$.
  Our strategy will be to bound the coefficients in $\epsilon$ of the central pressure, where $r=0$.
  We will then demonstrate that all coefficients in $\epsilon$ for $r > 0$ are less than the coefficients at $r=0$.
  From \citet[][Equation~(10.52), ignore their typo]{schutz2009first}
  \begin{align}
    \mathcal{P}(r) = \rho \frac{\sqrt{1 - 2\epsilon r^2/3R^2} - \sqrt{1 - 2\epsilon/3}}{3\sqrt{1 - 2\epsilon/3} - \sqrt{1 - 2\epsilon r^2/3R^2}}, \label{eqn:interior_sphere_pressure}
  \end{align}
  where we have substituted the equality in Equation~(\ref{eqn:defn_incompressible_epsilon}).
  Note that $\mathcal{P} = 0$ at $r = R$, so Assumption 2 is satisfied at the outer edge.
  Now consider $0 \leqslant r < R$ and define
  \begin{align}
    x &\equiv \sqrt{1 - 2\epsilon/3} \\
    y &\equiv \sqrt{1 - 2\epsilon r^2/3R^2},
  \end{align}
  so that
  \begin{align}
    \frac{\mathcal{P}(r)}{\rho} &= \frac{y - x}{3x - y} \\
    & = -1 + \frac{2}{3}\frac{1}{1 - y/3x}.
  \end{align}
  Note that $y/3x < 1$ on the interior of the object, so we may write the uniformly convergent series
  \begin{align}
    \frac{\mathcal{P}(r)}{\rho} = -1 + \frac{2}{3}\sum_{n=0}^\infty \left(\frac{y}{3x}\right)^n.
  \end{align}
  We now consider the expression on the rhs.
  Substitution of $x$ and $y$ gives
  \begin{align}
    -1 + \frac{2}{3}\sum_{n=0}^\infty \left[1 - \frac{2\epsilon}{3}\left(\frac{r}{R}\right)^2\right]^{n/2}\left[1 - \frac{2\epsilon}{3}\right]^{-n/2}\frac{1}{3^n}.
  \end{align}
 
  \subsection{Central Pressure Coefficient Bound}
  Setting $r=0$ gives
  \begin{align}
    \frac{\mathcal{P}(0)}{\rho} = -1 + \frac{2}{3}\sum_{n=0}^\infty \left[1 - \frac{2\epsilon}{3}\right]^{-n/2}\frac{1}{3^n}.
  \end{align}
  This expression satisfies the requirements for a uniformly convergent generalized binomial expansion:
  \begin{align}
    -1 + \frac{2}{3}\sum_{n=0}^\infty \sum_{k=0}^\infty \epsilon^k \binom{n/2 + k - 1}{n/2 - 1}\left(\frac{2}{3}\right)^k \frac{1}{3^n}.
  \end{align}
  Since both sums are uniformly convergent, we commute terms:
  \begin{align}
    -1 + \frac{2}{3}\sum_{k=0}^\infty \epsilon^k \left[ \sum_{n=0}^\infty \binom{n/2 + k - 1}{n/2 - 1}\left(\frac{2}{3}\right)^k \frac{1}{3^n}\right].
  \end{align}
  By definition of the binomial coefficients,
  \begin{align}
    \binom{n/2 + k - 1}{n/2 - 1} = \frac{1}{2^k k!}\prod_{j=0}^{k-1}\left(n + 2j\right),
  \end{align}
  we have that
  \begin{align}
        -1 + \sum_{k=0}^\infty \epsilon^k \left[ \sum_{n=0}^\infty \frac{2}{3^{(n+k+1)}k!}\prod_{j=0}^{k-1}\left(n + 2j\right)\right].
  \end{align}

  To demonstrate that $G_{ss}^{(k)}(0) \leqslant 1$, we establish that $c_k$, defined as
  \begin{align}
    c_k \equiv \sum_{n=0}^\infty c_n(k) \equiv \sum_{n=0}^\infty \frac{2}{3^{(n+k+1)}k!}\prod_{j=0}^{k-1}\left(n + 2j\right), \label{eqn:c_k}
  \end{align}
  monotonically decreases in $k$.
  First, note that all $c_n(k)$ are positive.
  We write the ratio of successive $c_n(k)$, at fixed $n$
  \begin{align}
    \frac{c_n(k+1)}{c_n(k)} = \frac{n + 2k}{3(k+1)}.
  \end{align}
  Note that the ratio is unity when $n = k + 3$.
  For terms with $n < k + 3$, individual $c_n(k)$ monotonically decrease in $k$ by the ratio test.
  We now partition off these terms:
  \begin{align}
    c_k \equiv \sum_{n=0}^{k+1} c_n(k) + \sum_{n=k+2}^\infty \frac{2}{3^{(n+k+1)}k!}\prod_{j=0}^{k-1}\left(n + 2j\right).
  \end{align}
  We need not assert that the finite sum itself is monotonically decreasing in $k$.
  We are only bookkeeping individual terms in the sum over $n$ that monotonically decrease in $k$.
  As $k$ increases, the number of such terms increases.
  It suffices to show that the summed tail contribution is monotonically decreasing in $k$.
  Once this is established, then all contributions to $c_k$ monotonically decrease in $k$. 
  This will establish that the $c_k$ monotonically decrease in $k$.
 
  We now proceed to show that the tail monotonically decreases in $k$.
  First, re-index the infinite series,
  \begin{align}
    c_k \equiv \sum_{n=0}^{k+1} c_n(k) + \sum_{m=0}^\infty \frac{2}{3^{(m+2k+3)}k!}\prod_{j=0}^{k-1}\left(m + k + 2 + 2j\right), \label{eqn:c_k-firstsplit}
  \end{align}
  and introduce notation for the coefficients of this infinite series
  \begin{align}
    \sum_{m=0}^\infty d_m(k) \equiv \sum_{m=0}^\infty \frac{2}{3^{(m+2k+3)}k!}\prod_{j=0}^{k-1}\left(m + k + 2 + 2j\right). \label{eqn:d-series}
  \end{align}
  Expanding the above product gives
  \begin{align}
    [m + (3k)][m+ (3k-2)]\dots[m + (3k-2k-2)],
  \end{align}
  where we have grouped each multiplicand as a binomial in $m$ and terms decrementing $k$.
  Note that there are $2^k$ grouped terms in its full expansion.
  For $d_n(k)$ where $m \geqslant 3k$, $2^k m^k$ is then an upper bound on the product.
  The tail of this series contribution to Equation~(\ref{eqn:c_k-firstsplit}) is then
  \begin{align}
    \sum_{m=3k}^\infty d_m(k) < \sum_{m=3k}^\infty \frac{2^{k+1}m^k}{3^{(m+2k+3)}k!} = \frac{2^{k+1}}{3^{(2k+3)}k!} \sum_{m=3k}^\infty \frac{m^k}{3^m}.
  \end{align}
  We bound the series with the integral
  \begin{align}
    \sum_{m=3k}^\infty \frac{m^k}{3^m} < \sum_{m=0}^\infty \frac{m^k}{3^m} < \frac{1}{(\ln 3)^{k+1}}\int_0^\infty p^k e^{-p}~\d p = \frac{k!}{(\ln 3)^{k+1}},
  \end{align}
  and obtain
  \begin{align}
    \sum_{m=3k}^\infty \frac{2^{k+1}m^k}{3^{(m+2k+3)}k!} < \frac{2^{k+1}}{3^{(2k+3)}(\ln 3)^{k+1}}.
  \end{align}
  This expression is also monotonically decreasing in $k$.

  The head of the series in Equation~(\ref{eqn:d-series}) contains terms $d_m(k)$, where $m < 3k$.
  We establish that these $d_m(k)$ monotonically decrease in $m$.
  The ratio of successive terms in $m$ is
  \begin{align}
    \frac{d_{m+1}(k)}{d_m(k)} = \frac{1}{3}\prod_{j=0}^{k-1}\left(1 + \frac{1}{m + k + 2 + 2j}\right).
  \end{align}
  Each term is largest when the denominator is smallest.
  This is achieved when $j=m=0$, giving an upper bound for the product
  \begin{align}
    \frac{d_{m+1}(k)}{d_m(k)} \leqslant \frac{1}{3}\left(1 + \frac{1}{k + 2}\right)^k < \frac{e}{3} < 1.
  \end{align}
  We conclude the terms monotonically decrease in $m$ for fixed $k$.
  The multiplicands in the head $d_m(k)$ are largest when $m = 0$, giving the following upper bound
  \begin{align}
    \sum_{m=0}^{3k-1} d_m(k) < 6k\frac{(3k)!!}{3^{2k+3}k!k!!}.
  \end{align}
  This expression is also monotonically decreasing in $k$. 

  We may (finally) write a monotonically decreasing bound for $c_k$
  \begin{align}
        c_k < \sum_{n=0}^{k+1} c_n(k) + \frac{2^{k+1}}{3^{(2k+3)}(\ln 3)^{k+1}} + 6k\frac{(3k)!!}{3^{2k+3}k!k!!} \label{eqn:c_k-bound}.
  \end{align}
  To show that all $c_k < 1$, it now suffices to explicitly compute the $k=0$ bound.
  Substitution of $k=0$ into Equation~(\ref{eqn:c_k-bound}) gives
  \begin{align}
    c_0 < \left[c_0(0) + c_1(0) + \frac{2}{3^3 \ln 3}\right] < \frac{26}{27} < 1.
  \end{align}
  Because $c_k \geqslant 0$ always, this completes the proof that $\left|G^{(k)}_{ss}(0)\right| \leqslant 1$ as required by Assumption 2.
  
  As a check on our work, central coefficients can be explicitly computed through parametric differentiation with respect to $\epsilon$ of Equation~(\ref{eqn:interior_sphere_pressure})
  \begin{align}
    c_k =  \lim_{\epsilon \to 0} \frac{1}{k!} \frac{\partial^k}{\partial\epsilon^k}\mathcal{P}(0).
  \end{align}
  For example, evaluation at $k=1$ gives
  \begin{align}
    c_1 = \frac{1}{6},
  \end{align}
  which agrees with Equation~(\ref{eqn:c_k}) evaluated at $k=1$.
  As an additional check, the bound has been verified numerically through $c_{60}$.
  
  \subsection{Peripheral Pressure Coefficient Bound}
  Now that we have established that the coefficients of the central pressure are bounded, we show that the coefficients for $r > 0$ remain bounded.
  This is done readily with series long division.
  We will show that each coefficient at $r > 0$ is decreased relative to its value at $r = 0$ and that coefficients remain bounded below by zero.
  
  To prepare for the algorithm, note that Equation~(\ref{eqn:interior_sphere_pressure}) can be written using uniformly convergent binomial expansions as
  \begin{align}
    \frac{\mathcal{P}(r)}{\rho} -1 = \left(1 + \sum_{j=1}^\infty a_j\epsilon^j\right)\Bigg/\left\{1 + \frac{1}{2}\sum_{k=1}^\infty a_k \epsilon^k\left[3 - \left(\frac{r}{R}\right)^{2k}\right]\right\}, \label{eqn:long-division-form}
  \end{align}
  where the $a_n$ are defined as
  \begin{align}
    a_n \equiv  (-1)^n\left(\frac{2}{3}\right)^n \binom{1/2}{n}.
  \end{align}
  The long division algorithm states that, if
  \begin{align}
    \sum_{k=0}^\infty c_k\epsilon^k \equiv \sum_{j=0}^\infty a_j\epsilon^j \Bigg/ \sum_{\ell=0}^\infty b_\ell\epsilon^\ell,
  \end{align}
  then
  \begin{align}
    c_0 &= \frac{a_0}{b_0}\\
    c_k &= \frac{1}{b_0}\left(a_k - \sum_{j=0}^{k-1}c_jb_{k-j} \right) \qquad (k \geqslant 1).
  \end{align}
  By inspection of Equation~(\ref{eqn:long-division-form}), we see that
  \begin{align}
    a_0 = b_0 = c_0 = 1.
  \end{align}
  Now, note that
  \begin{align}
    \binom{1/2}{j} \propto (-1)^{j+1} \qquad (j \geqslant 1).
  \end{align}
  By inspection of Equation~(\ref{eqn:long-division-form}), we may now write 
  \begin{align}
    c_k(r) = -\left|a_k\right| + \frac{1}{2}\sum_{j=0}^{k-1}c_j(r)\left|a_{k-j}\right|\left[3 - \left(\frac{r}{R}\right)^{2(k-j)}\right].
  \end{align}
  Pull off the first term of the sum and group the $a_k$ terms
  \begin{align}
    \begin{split}
      c_k(r) =& \left|a_k\right|\left\{\frac{1}{2}c_0(r)\left[3 - \left(\frac{r}{R}\right)^{2k}\right] - 1\right\} \\
      &+ \frac{1}{2}\sum_{j=1}^{k-1}c_j(r)\left|a_{k-j}\right|\left[3 - \left(\frac{r}{R}\right)^{2(k-j)}\right].
    \end{split} \label{eqn:periphery}
  \end{align}
  It is now clear that $c_k(r) \geqslant 0 \implies c_{k+1}(r) \geqslant 0$, provided that
  \begin{align}
    \frac{1}{2}c_0(r)\left[3 - \left(\frac{r}{R}\right)^{2k}\right] \geqslant 1.
  \end{align}
  Using that $c_0 = 1$, it follows that the all $c_k(r)$ are positive if
  \begin{align}
    \left(\frac{r}{R}\right)^{2k} \leqslant 1,
  \end{align}
  which is always true interior to the sphere.
  By inspection of Equation~(\ref{eqn:periphery}), we can now conclude that each $c_k(r)$ decreases monotonically in $r$.
  Since we have already bounded the central coefficients, it follows that $\left|G_{ss}^{(k)}(r)\right| < 1$ for all $k$ within the constant-density sphere, as was to be shown.  

\end{appendix}

\bibliographystyle{aasjournal}
\bibliography{thesis}

\end{document}